\documentclass[AMA,Times1COL]{WileyNJDv5} %STIX1COL,STIX2COL,STIXSMALL

\usepackage{subfiles} % Best loaded last in the preamble
\usepackage{multirow}
\usepackage{amssymb}
\articletype{Article Type}%

\received{Date Month Year}
\revised{Date Month Year}
\accepted{Date Month Year}
\journal{Journal}
\volume{00}
\copyyear{2023}
\startpage{1}

\begin{document}

\title{FinRL Contests: Benchmarking Data-driven Financial Reinforcement Learning Agents}

\author[1]{Keyi Wang}

\author[2]{Nikolaus Holzer}

\author[2]{Ziyi Xia}

\author[4]{Yupeng Cao}

\author[1,5]{Jiechao Gao}

\author[1,3]{\\Anwar Walid}

\author[6]{Kairong Xiao}

\author[1,3]{Xiao-Yang Liu Yanglet}

\authormark{Wang \textsc{et al.}}
\titlemark{FinRL Contests: Benchmarking Data-driven Financial Reinforcement Learning Agents}

\address[1]{\orgdiv{SecureFinAI Lab}, \orgname{Columbia University}, \orgaddress{\state{New York}, \country{USA}}}

\address[2]{\orgdiv{Computer Science}, \orgname{Columbia University}, \orgaddress{\state{New York}, \country{USA}}}

\address[3]{\orgdiv{Electrical Engineering}, \orgname{Columbia University}, \orgaddress{\state{New York}, \country{USA}}}

\address[4]{\orgdiv{Electrical and Computer Engineering}, \orgname{Stevens Institute of Technology}, \orgaddress{\state{New Jersey}, \country{USA}}}

\address[5]{\orgdiv{Center for SDGC}, \orgname{Stanford University}, \orgaddress{\state{California}, \country{USA}}}

\address[6]{\orgdiv{Columbia Business School}, \orgname{Columbia University}, \orgaddress{\state{New York}, \country{USA}}}

\corres{Corresponding author Xiao-Yang Liu Yanglet \email{
XL2427@columbia.edu}}

% \presentaddress{This is sample for present address text this is sample for present address text.}

%\fundingInfo{Text}
%\JELinfo{ejlje}

\abstract[Abstract]{Financial reinforcement learning (FinRL) is now a practical paradigm for financial engineering. However, applying RL strategies to real-world trading tasks remains a challenge for individuals, as it is error-prone and engineering-heavy. The non-stationarity of financial data, low signal-to-noise ratios, and various market frictions require deep accumulations. Although numerous FinRL methods have been developed for tasks such as stock/crypto trading and portfolio management, the lack of standardized task definitions, real-time high-quality datasets, close-to-real market environments, and robust baselines has hindered consistent reproduction in both open-source community and FinTech industry. To bridge this gap, we organized a series of FinRL Contests from 2023 to 2025, covering a diverse range of financial tasks such as stock trading, order execution, crypto trading, and the use of large language model (LLM)-engineered signals. These contests attracted 200+ participants from 100+ institutions over 20+ countries. To encourage participations, we provided starter kits featuring GPU-optimized parallel market environments, ensemble learning, and comprehensive instructions. In this paper, we summarize these benchmarking efforts, detailing task formulations, data curation pipelines, environment implementations, evaluation protocols, participant performance, and organizational insights. It guides our follow-up FinRL contests, and also provides a reference for FinAI contests alike.}


\jnlcitation{\cname{%
\author{Keyi Wang},
\author{Lauritzen P},
\author{Erath C}, and
\author{Mittal R}}.
\ctitle{On simplifying ‘incremental remap’-based transport schemes.} \cjournal{\it J Comput Phys.} \cvol{2021;00(00):1--18}.}

\maketitle

\renewcommand\thefootnote{}
% \footnotetext{\textbf{Abbreviations:} ANA, anti-nuclear antibodies; APC, antigen-presenting cells; IRF, interferon regulatory factor.}

\renewcommand\thefootnote{\fnsymbol{footnote}}
\setcounter{footnote}{1}

\section{Introduction}\label{sec1}

% \documentclass[../main.tex]{subfiles}
% \graphicspath{{\subfix{../fig/}}}
% \begin{document}

% Applications of FinRL

% Motivation.
% \begin{itemize}
%     \item challenges of financial markets and financial data
%     \item Reproduction: benchmarks and validation
% \end{itemize}

% Report of FinRL Contests 2023-2025
% \begin{itemize}
%     \item Overview of FinRL contests.
%     \item Benchmarking performance.
%     \item Open-source (repo, documentation website)
% \end{itemize}

% No need to discuss challenges of data here
% Require reproducible benchmark
% Policy instability and sampling bottleneck
Financial reinforcement learning (FinRL)~\cite{liu2020finrl, liu2021finrl, liu2018practical, li2021finrl, liu2022finrl, liu2024dynamic} is an interdisciplinary field that applies reinforcement learning algorithms to financial tasks, such as algorithmic trading, portfolio management, option pricing, hedging, and market making \cite{charpentier2021reinforcement, fischer2018reinforcement, hambly2023recent, sun2023reinforcement, bai2025review}. FinRL aims to train trading agents to interact with dynamic financial environments and to optimize their financial decisions autonomously. As FinRL methods continue to advance, two practical challenges have become increasingly apparent. One is policy instability, where small changes in training settings or market environments can lead to large variation in performance. The other is the sampling bottleneck, as collecting high-quality trajectories is often expensive and limited by access to financial data. Additionally, it remains difficult to compare the performance of these RL strategies due to inconsistencies in task definitions, dataset choices, environment implementations, and baselines. These challenges make it difficult to obtain reliable and consistent results, emphasizing the need for reproducible benchmarks that enable fair and systematic evaluation across tasks.

% However, applying RL to financial markets faces several unique challenges. First, market data is dynamic and non-stationary, which is different from the static datasets commonly used in machine learning. Market behavior changes over time due to economic trends, investor sentiment, and regulatory updates, making it difficult to model and predict. Second, financial data has a low signal-to-noise ratio, which makes it harder to extract alpha signals and increases the risk of overfitting. Third, real-world trading involves market frictions, such as transaction costs and liquidity risks, which make FinRL environments more complex than standard RL settings such as Atari~\cite{mnih2013playing} or MuJoCo~\cite{todorov2012mujoco}.

 %However, it remains difficult to compare the performance of these RL strategies due to inconsistencies in task definitions, dataset choices, environment implementations, and baselines. 
 FinRL-Meta~\cite{liu2022finrl, liu2024dynamic} partially addressed this issue by transforming financial data into gym-style market environments, which built an automated data curation pipeline to deal with dynamic market data. Such a framework enables different FinRL algorithms to be implemented and evaluated under standardized conditions across various tasks. Still, there is no widely accepted benchmark for evaluating RL agents in financial applications~\cite{sun2023reinforcement}, which hinders reproducibility and makes it difficult for the community to compare results and build on prior works. There is a strong need for an open and collaborative platform where researchers can systematically evaluate their methods under realistic financial conditions. To fill this gap, we initiated the FinRL Contests as a community-driven effort to promote reproducibility, transparency, and benchmarking in financial reinforcement learning.

From 2023 to 2025, we have organized a series of FinRL contests to support reproducible experimentation and systematic benchmarking. These contests cover diverse tasks such as stock trading, order execution, crypto trading, and LLM-engineered signals for FinRL. Excluding the ongoing FinAI Contest 2025, the FinRL contests attracted 200+ participants from 100+ institutions over 20+ countries. They formed 46, 21, and 85 teams across 2023 to 2025, respectively. To ensure transparency and reproducibility, all participant teams were encouraged to open-source the solutions. To support the development of solutions for the FinRL Contest tasks, we provide starter kits for participants, including tutorials, market data, GPU-optimized parallel market environments, and baseline methods. The full list of starter kits and contest websites from 2023 to 2025 is summarized in Table~\ref{tab:kits}. In addition, we maintain a continuously updated documentation, serving as a detailed guide and resource hub for participants. This documentation includes a detailed FinRL introduction, task descriptions, and explanations of baseline solutions. 

In this paper, we summarize the FinRL Contests 2023-2025 that are our benchmarking efforts for financial reinforcement learning. We detail the formulation of financial tasks as Markov Decision Processes (MDPs), the design of an automated data curation pipeline, the implementation of GPU-optimized parallel market environments, and the establishment of standardized evaluation protocols. In addition, we provide a summary of participant performance and the organizational experience gained from running these contests.

The remainder of this paper is organized as follows. Section 2 reviews related works. Section 3 describes tasks. Section 4 describes dynamic market data in FinRL. Section 5 describes the benchmarking pipeline. Section 6 shows performance. Section 7 summarizes organizational aspects. Finally, we conclude this paper in Section 8.

% \begin{figure}
%     \centering
%     \includegraphics[width=0.6\linewidth]{fig/ML_DRL_comparison.png}
%     \caption{Comparison between conventional ML approach and RL in Finance.}
%     \label{fig:ml_drl_compare}
% \end{figure}
\begin{table}[t]
\centering
\begin{tabular}{|c|l|l|}
\hline
\textbf{FinRL Contests} & \textbf{Starter Kit} & \textbf{Website} \\ \hline
       2025 @ IEEE CSCloud*           &      \href{https://github.com/Open-Finance-Lab/FinAI_Contest_2025}{Github Repo 2025}    &    \href{https://open-finance-lab.github.io/FinAI_Contest_2025/}{Website 2025}    \\ \hline
       2025 @ IEEE IDS           &      \href{https://github.com/Open-Finance-Lab/FinRL_Contest_2025}{Github Repo 2025}    &    \href{https://open-finance-lab.github.io/FinRL_Contest_2025/}{Website 2025}    \\ \hline
    2024 @ ACM 2024            &  \href{https://github.com/Open-Finance-Lab/FinRL_Contest_2024}{Github Repo 2024}  & \href{https://open-finance-lab.github.io/finrl-contest-2024.github.io/}{Website 2024}\\ \hline
    2023 @ ACM 2023             &  \href{https://github.com/Open-Finance-Lab/FinRL_Contest_2023}{{Github Repo 2023} } & \href{https://open-finance-lab.github.io/finrl-contest.github.io/}{Website 2023}\\ \hline
    Documentation & \multicolumn{2}{|c|}{\href{https://finrl-contest.readthedocs.io/en/latest/}{FinAI Contests documentation}}  \\ \hline
\end{tabular}
\caption{Links to the starter kits and documentation provided for FinRL Contests 2023-2025. * The FinAI Contest 2025 @ IEEE CSCloud is ongoing.}
\label{tab:kits}
\end{table}

% \end{document}

\section{Related Work}\label{sec2}

% \documentclass[../main.tex]{subfiles}
% \graphicspath{{\subfix{../fig/}}}
% \begin{document}

\subsection{Applications of Financial Reinforcement Learning}

% cite survey paper

Financial markets are inherently complex, dynamic, and high-dimensional, making deep reinforcement learning (DRL) a compelling approach for sequential decision-making~\cite{hambly2023recent}. DRL has been successfully applied to a variety of financial tasks~\cite{fischer2018reinforcement}, including algorithmic trading, portfolio optimization, and option pricing, where traditional rule-based or supervised methods often fall short in capturing long-term rewards under uncertainty. Beyond conventional reward-driven learning, recent advances in preference-based reinforcement learning offer alternative supervision signals. For example, Direct Preference Optimization (DPO)~\cite{rafailov2023direct} formulates learning objectives based on pairwise comparisons between trajectories, enabling policy improvement in settings where explicit reward design is difficult or noisy. Such methods hold promise for financial applications that involve subjective goals, human feedback, or implicit utility functions, and may further enhance the robustness and flexibility of DRL in real-world markets.

\subsection{Methods of Financial Reinforcement Learning}

% Sutton's book
% cite survey paper, website, book

% Deep reinforcement learning (DRL) algorithms are commonly categorized into value-based, policy-based, and actor-critic methods, each offering complementary strengths for decision-making. Value-based methods, such as Q-learning~\cite{watkins1992q} and its deep extensions (e.g., DQN, Double DQN, Dueling DQN~\cite{SpinningUp2018}), estimate action-value functions to guide policy learning. These methods have shown success in discrete action spaces but face challenges when applied to continuous control problems often found in finance. Policy-based approaches overcome this limitation by directly optimizing parameterized policies via policy gradients~\cite{Sutton00policygradient}, enabling applications in high-dimensional and continuous domains, such as portfolio allocation and dynamic pricing. Actor-critic methods combine value estimation and policy learning by jointly training an actor (policy) and a critic (value function), offering improved stability and sample efficiency. Algorithms like PPO, DDPG, SAC, and TD3~\cite{SpinningUp2018} are widely adopted in financial RL due to their scalability and robustness in volatile environments. These algorithmic foundations have enabled a growing wave of research focused on deploying DRL for complex financial tasks. 

A wide range of reinforcement learning methods \cite{charpentier2021reinforcement, fischer2018reinforcement} have been explored for financial decision-making. Among these, algorithms based on policy gradients and actor-critic architectures \cite{SpinningUp2018} have received particular attention due to their effectiveness in handling continuous control, high-dimensional action spaces, and non-stationary dynamics. Notable examples include proximal policy optimization (PPO), deep deterministic policy gradient (DDPG), soft actor-critic (SAC), and twin-delayed DDPG (TD3), which are frequently adopted for their robustness and sample efficiency. Ying et al. \cite{ying2022towards} proposed the CVaR Proximal Policy Optimization (CPPO) algorithm, which integrates Conditional Value-at-Risk (CVaR) into its objective based on PPO.

%uses Conditional Value-at-Risk (CVaR) to assess and manage risk during the learning process.

Building on these algorithmic foundations, many studies have explored DRL applications in quantitative finance. For example, Deng et al.~\cite{deng2016deep} propose a recurrent deep reinforcement learning framework that jointly learns market representations and trading policies. Their model demonstrates robust performance across stock and commodity markets by integrating deep feature extraction with sequential decision-making. Avramelou et al.~\cite{avramelou2024deep} propose a multi-modal embedding approach to integrate both price and sentiment data in DRL-based trading agents. Their method improves trading performance while enabling dynamic adjustment of modality importance without retraining. To address the practical challenges of applying DRL in quantitative trading, Li et al.~\cite{li2021finrl} propose FinRL-Podracer, a scalable cloud-based framework for efficient training and deployment of DRL agents. The system integrates high-performance GPU optimization with automated training workflows, significantly improving both trading performance and development efficiency. While recent studies have advanced DRL applications across various financial domains, systematic evaluation and comparison of different approaches remain an important direction, motivating further efforts toward benchmarking financial reinforcement learning.

% DRL algos for Finance and finance benchmark

\subsection{Benchmarks of Financial Reinforcement Learning}

Recent research works have applied deep reinforcement learning (DRL) to quantitative finance~\cite{yang2020deep, zhang2020deep, ardon2021towards, amrouni2021abides, coletta2021towards} by developing customized market environments. Although these open-source libraries provide useful environments and datasets from sources like Yahoo Finance and WRDS, there is still a lack of standardized benchmarks for systematic evaluation. FinRL~\cite{liu2020finrl, liu2021finrl,liu2018practical} offers a full pipeline with environments for stock trading, portfolio allocation, and crypto trading; however, these environments are increasingly insufficient to meet the growing demands of the community. To address the challenges of data quality, survivorship bias, and model overfitting, Liu et al.~\cite{liu2022finrl, liu2024dynamic} further introduced FinRL-Meta. FinRL-Meta follows a DataOps paradigm to automatically collect dynamic datasets and construct hundreds of gym-style market environments, reproduces popular DRL trading papers to facilitate research reproducibility, and supports cloud-based deployment and visualization for community-driven performance evaluation. Despite these advances, establishing a unified, large-scale, and competitive benchmark for data-driven financial reinforcement learning remains an open direction, motivating the development of FinRL Contests.

% \end{document}

% \subsection{Review of FinRL Applications}

% \subsection{Deep Reinforcement Learning Algorithms}

% in finance

% \subsection{Benchmarking FinRL Performance}

\section{Tasks}
% \documentclass[../main.tex]{subfiles}
% \graphicspath{{\subfix{../fig/}}}
% \begin{document}

Based on projects FinRL \cite{liu2020finrl, liu2021finrl, liu2018practical, li2021finrl} and FinRL-Meta \cite{liu2022finrl, liu2024dynamic}, and feedback from our open-source community, we select several stable and representative trading tasks for the FinRL Contests. In this section, we describe the tasks in the language of Markov Decision Processes (MDPs), discuss the parallelism in the estimate of policy gradients, and describe the specific tasks featured in the FinRL Contests.

\subsection{Tasks in FinRL Contests 2023-2025}

% Accumulation: cite papers, attract others, add criteria for selecting tasks. Give more reasons from community, relate to the related works. 增加分量

\begin{figure}
    \centering
    \includegraphics[width=\linewidth]{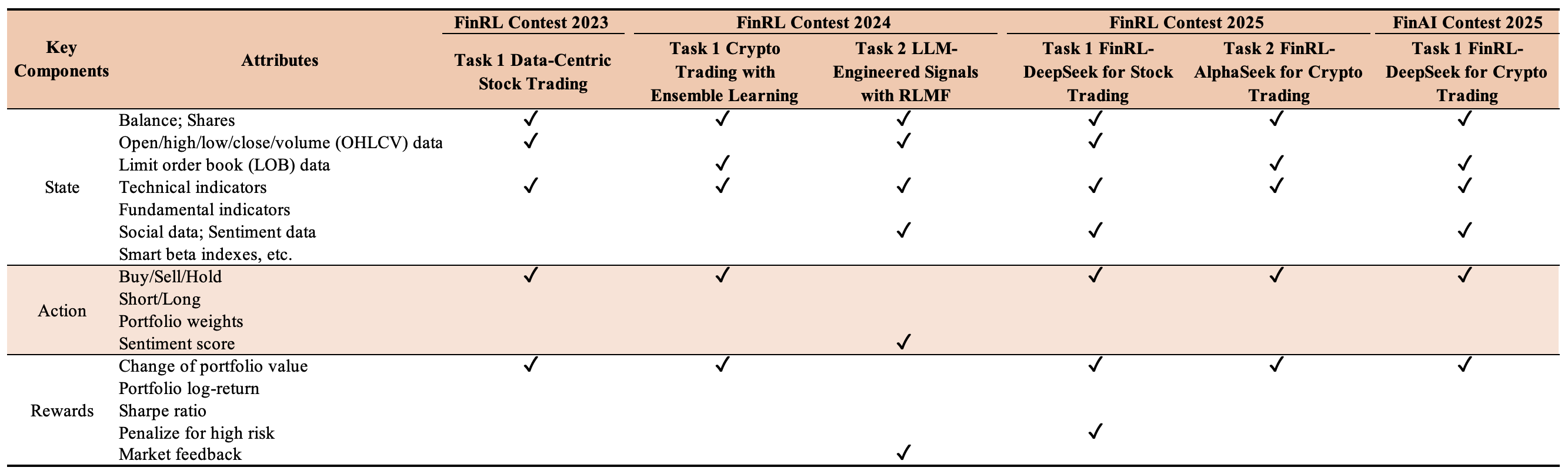}
    \caption{Tasks in FinRL Contests 2023-2025.}
    \label{fig:tasks}
\end{figure}

We summarize trading tasks of FinRL Contests 2023-2025 in Fig. \ref{fig:tasks}, including daily stock trading with OHLCV data and second-level crypto trading with LOB data. The stock trading tasks include:
\begin{itemize}
    \item \textbf{Data-Centric Stock Trading @ FinRL Contest 2023} \cite{liu2020finrl, liu2021finrl,liu2018practical}. It is to train a stock trading agent by applying novel data and feature engineering based on market data. It is a multi-asset trading task for 30 constituent stocks in the Dow Jones Index. The task has been clearly defined and well studied in FinRL \cite{liu2020finrl, liu2021finrl,liu2018practical} and FinRL-Meta \cite{liu2022finrl, liu2024dynamic}. The task is stable and reproducible, making it ideal for benchmarking in a competition setting. In addition, this task focuses on data and feature engineering, encouraging participants to tackle the challenge of dynamic market data. This aligns with the community’s ongoing interest in exploring dynamic financial data and developing stable multi-asset environments. 
    % \item \textbf{FinRL Contest 2023 Real Time Order Execution}. This sub-task focuses on building a lightweight algorithmic trading agent in a fair environment.
    \item \textbf{LLM-Engineered Signals with RLMF @ FinRL Contest 2024} \cite{liu2023fingpt, nie2024llmsurevey}. It is to develop LLMs that can generate actionable trading signals from financial news by using reinforcement learning from market feedback (RLMF). LLMs have been applied in many financial tasks \cite{nie2024llmsurevey}, but general-purpose LLMs trained on Internet data cannot capture the intrinsic dynamics of financial markets. To align LLMs with financial markets, this task encourages adapting LLMs using Reinforcement Learning from Market Feedback, as a counterpart to Reinforcement Learning from Human Feedback (RLHF) \cite{christiano2017rlhf}. RLMF utilizes feedback from the financial market as reward signals, enabling LLMs to learn from financial market behaviors. This task was selected for its novelty and potential to push the frontier of LLM-enhanced financial decision-making. It aligns with the growing interest in the FinRL and FinLLM communities in combining FinRL and LLMs.
    %By incorporating market feedback in the fine-tuning process, LLMs can learn from and adapt to financial market behaviors.
    \item \textbf{FinRL-DeepSeek for Stock Trading @ FinRL Contest 2025} \cite{liu2023fingpt, deepseek2025mostapha}. It is to develop trading agents by combining FinRL and LLM-engineered signals. Unstructured financial texts, such as news and SEC filings, contain valuable information about market sentiment, risk, and events. LLMs, such as DeepSeek-V3, have been applied to extract signals from financial documents \cite{nie2024llmsurevey}, which are then integrated into FinRL. This task was selected because it reflects real-world decision-making, where financial analysts and traders rely on both structured market data and unstructured financial text. It also aligns with the growing interest in the FinRL and FinLLM communities in LLM-engineered signals and multimodal data exploration.
\end{itemize}

Different from stock markets, crypto trading faces greater volatility due to drastic price fluctuations and market sentiment shifts. crypto markets operate 24/7, demanding adaptable strategies in a continuous trading environment without traditional market open and close cycles. It is more challenging to develop robust and profitable trading strategies. To reflect these unique challenges, we include crypto trading tasks in the FinRL Contests. To differentiate from the multi-asset stock trading tasks above, the crypto trading tasks are designed to trade single-asset Bitcoin at the second level. The crypto trading tasks include:
\begin{itemize}
    \item \textbf{crypto Trading with Ensemble Learning @ FinRL Contest 2024 } \cite{yang2020deep} is to train a trading agent for Bitcoin by using ensemble methods. RL policies are unstable, whose performance is sensitive to hyperparameters, market noise, and random seeds. Policy instability can also come from value function approximation errors. These issues are amplified in crypto markets with high volatility. Ensemble methods have shown effectiveness in mitigating policy instability \cite{yang2020deep}. This task extends these ideas to the crypto trading tasks. In addition, collecting high-quality financial trajectories is often expensive and limited by data access. It causes the sampling bottleneck, especially during training multiple FinRL agents for an ensemble. Therefore, this task encourages participants to employ novel ensemble methods and utilize the provided GPU-optimized parallel environments to improve sampling speed. It also aligns with the need in the financial industry for robust and efficient trading strategies in crypto markets. 
    \item \textbf{FinRL-AlphaSeek for Crypto Trading @ FinRL Contest 2025 } \cite{yang2020deep} extends the previous task of crypto trading by introducing an additional focus on factor mining. Factors such as momentum play a critical role in driving trading decisions, enabling traders to design efficient, data-driven strategies. This task encourages participants to develop a two-stage pipeline: 1) factor engineering and selection and 2) ensemble learning to build robust trading agents. It aligns with the industry need for effective alpha and beta signal extraction in crypto markets.
    \item \textbf{FinRL-DeepSeek for Crypto Trading @ FinAI Contest 2025 } \cite{yang2020deep} extends the previous two tasks of crypto trading by introducing LLM-engineered signals. crypto markets are highly sensitive to news headlines, tweets, regulatory shifts, and viral narratives. The timely interpretation of market sentiment is critical for crypto trading. LLMs can extract actionable signals from financial news, tweets, and filings. In this task, we encourage participants to explore LLM-engineered signals and integrate them into a FinRL trading agent for crypto trading.
\end{itemize}
% \textbf{FinRL Contest 2024 crypto Trading with Ensemble Learning} and \textbf{FinRL Contest 2025 FinRL-AlphaSeek for Crypto Trading}. These tasks are to train a trading agent for Bitcoin. Participants are encouraged to explore different factor mining approaches and ensemble methods. 

% Decentralized Finance (DeFi) is reshaping the crypto economy by enabling peer-to-peer trading, lending, and liquidity provision without intermediaries. In the ongoing FinAI Contest 2025, we also include a task about DeFi liquidity provisioning:
% \begin{itemize}
%     \item \textbf{FinRL-DeFi @ FinAI Contest 2025}\cite{xu2025defi}. As a core component of DeFi, the automated market makers (AMMs) act as liquidity providers (LPs) and replace order books with liquidity pools. However, liquidity provision is complex and risky. For example, impermanent loss can occur for LPs when the price of assets in a liquidity pool diverges from their initial value. LPs must actively manage price ranges, balance transaction fees, and mitigate impermanent loss. In this task, we challenge participants to develop FinRL agents that act as LPs, dynamically adjusting their liquidity positions in response to market conditions.
% \end{itemize}

To address the sampling bottleneck of the training stage, we develop massively parallel market environments on GPUs. We encourage contestants to use ensemble methods to mitigate policy instability and improve models' robustness. These will be described in the following section.

\subsection{MDP Formulation}\label{sec:mdp}
% hightlight the future information leakage
% data -> feature (state, action, reward) -> partition --> step reward reset

%\subsection{Task Description}\label{sec:mdp}
We formulate stock and crypto trading tasks as Markov Decision Processes (MDPs)~\cite{liu2024dynamic}.
\begin{itemize}
    \item \textbf{State} $\mathbf{s_t} = [b_t, \mathbf{p_t}, \mathbf{h_t}, \mathbf{f_t}] \in \mathbb{R}^{K(I+2)+1}$ represents market conditions at time $t$. $b_t \in \mathbb{R}_+$ is the account balance. $\mathbf{p_t} \in \mathbb{R}_+^{K}$ is the prices of stocks or cryptocurrencies, where $K$ is the number of assets. $\mathbf{h_t} \in \mathbb{R}_+^{K}$ is the holding positions. $\mathbf{f_t} \in \mathbb{R}^{KI}$ is the feature vector, where there are $I$ features for each asset. 
    %, such as Moving Average Convergence Divergence (MACD) and sentiment score derived from the news.
    \item \textbf{Action} $\mathbf{a_t} \in \mathbb{R}^{K}$ represents trading actions at time $t$, such that $\mathbf{h}_{t+1} = \max(\mathbf{h_t} + \mathbf{a_t}, 0)$ (making $\mathbf{h}_{t+1}$ nonnegative). An entry $a_t^i > 0, i = 1,\ldots, K$ indicates buying $a_t^i$ shares of assets $i$, while $a_t^i < 0$ indicates selling and $a_t^i = 0$ indicates holding the current position. Each entry of $\mathbf{h_{t+1}}$ is nonnegative, which means the agent cannot sell more shares than it currently holds.

    \item \textbf{State-transition function} $\delta(\mathbf{s}_{t+1}|\mathbf{s}_t,\mathbf{a}_t)$ defines the probability distribution over the next state $\mathbf{s}_{t+1}$ given the current state $\mathbf{s}_{t}$ and action $\mathbf{a}_{t}$. Transitions are governed by real-world asset price movements and trading outcomes. In simulation with historical market data, the transitions may be deterministic during training. In real-time trading, the state transition function is stochastic.
    \item \textbf{Reward} $r_{t+1}=R(\mathbf{s_t}, \mathbf{a_t}, \mathbf{s_{t+1}})$ is an incentive signal to encourage or discourage the trading agent to perform action $\mathbf{a_t}$ at state $\mathbf{s_t}$. We use $r_{t+1}$ instead of $r_t$ to reflect the delayed reward in financial markets. It emphasizes that the reward $r_{t+1}$ is determined at the next state $\mathbf{s}_{t+1}$ after taking action $\mathbf{a}_t$. The reward can be defined as the change in the total asset value, i.e., $R(\mathbf{s_t}, \mathbf{a_t}, \mathbf{s_{t+1}}) = v_{t+1} - v_{t}$, where $v_t = b_t + \mathbf{p_t}^T\mathbf{h_t}$ is the total asset value at time $t$. The reward can also be penalized for high-risk signals. 
    \item \textbf{Discount factor} $\gamma \in (0,1)$ is used to evaluate discounted expected return at time t. The discount factor $\gamma$ determines the present value of future rewards.
\end{itemize}
Note that a policy $\pi( \cdot  \mid  \mathbf{s}_t)$ is a probability distribution over actions at state $\mathbf{s}_t$. It determines the likelihood of each possible trading action given the current market conditions. The trading agent’s objective is to learn a policy $\pi$ that maximizes cumulative positive rewards while minimizing losses over time.

In the state $\mathbf{s}_t$, the holding positions $\mathbf{h}_t$ are typically constrained to be nonnegative in long-only trading settings, where the agent is not allowed to short assets. The feature vector $\mathbf{f}_t$ can include market indicators (e.g., moving average convergence divergence), factors learned by neural networks, and LLM-engineered signals from financial news or regulatory filings. The action $\mathbf{a}_t$ can be discrete or continuous, depending on the task. It may support long-only or long-short strategies. The reward function $R$ can use the Sharpe ratio in addition to the change in asset value. It can be further adjusted by risk or market sentiment to better reflect real-world trading scenarios. 

For example, a trading task is to develop a trading agent for all 30 constituent stocks in the Dow Jones Index with an initial investment of $\$1$ million. In this setting, the number of assets is $K = 30$. The state includes the account balance of $b_0=100,000$, prices of the 30 stocks, holding positions of the 30 stocks, and $I=4$ technical indicators for each stock. The state space has a dimension $K(I+2)+1=181$. The action space has dimension $K=30$. An entry, for example, $a_t^1 = 5$ means buying $5$ shares of the first stock in the portfolio. The reward function is the change in the total asset value, with $v_0 = 100,000$. Taking the first stock as an example, assume at time $t$, the account balance is $b_t = 100,000$, the price of the first stock is $p_t^1=200$, and the shares of the first stock are $h_t^1=10$. The agent buys 5 shares of the first stock ($a_t^1 = 5$) and the price increases to $p_{t+1}^1=201$ at time $t+1$. Assuming holding all other 29 stocks, the state is updated where $b_{t+1} = 99,000$, $p_{t+1}^1=201$, and $h_{t+1}=15$. The return $r_t = v_{t+1} - v_t = (b_{t+1} + \mathbf{p}_{t+1}^T\mathbf{h}_{t+1}) - (b_t + \mathbf{p}_{t}^T\mathbf{h}_{t}) = b_{t+1} - b_t + p_{t+1}^1h_{t+1}^1 - p_{t}^1h_{t}^1 = -1000 + 201\cdot15 - 200 \cdot 10 = 15$. In this example,  the agent only operates on the first stock. 

\subsection{Policy Gradient Estimate and Its Parallelism}

In the FinRL tasks, the policy $\pi_\theta$ is typically a neural network (e.g., transformer network or diffusion model), where $\theta$ denotes the learnable parameters. A trajectory, $\tau = (\mathbf{s}_0, \mathbf{a}_0, r_1, \mathbf{s}_1, \mathbf{a}_1, r_2, \mathbf{s}_2, \mathbf{a}_2,\ldots, \mathbf{s}_T)$, is a sequence of trading actions and states observed over a period, where $\mathbf{s}_0$ is the initial state and $T$ is the number of steps. The probability of the trajectory $\tau$ is 
\begin{equation}
P(\tau  \mid  \pi_{\theta}) = \rho_0(\mathbf{s}_0) \prod_{t=0}^{T-1} \pi_{\theta}(\mathbf{a}_t \mid \mathbf{s}_t) P(\mathbf{s}_{t+1}  \mid \mathbf{s}_t, \mathbf{a}_t),
\end{equation}
where $\rho_0(s_0)$ is the initial state distribution.
The financial results are quantified as rewards along $\tau$ to calculate the cumulative return $R(\tau) = \sum_{t=1}^{T} \gamma^{t-1} r_t$. The goal is to learn a policy $\pi_{\theta}$ that maximizes the following expected return
\begin{equation}
    J(\theta) = \mathbb{E}_{\tau \sim \pi_{\theta}}[R(\tau)] = \int_{\tau} P(\tau \mid \pi_{\theta})R(\tau),
\end{equation}
%where $R(\tau) = $ is the (discounted) cumulative return along the trajectory $\tau$. 
The gradient of $J(\theta)$ with respect to $\theta$ is ~\cite{peters2010gradient}
\begin{equation}
\label{eq:gradient_J}
    \nabla J(\theta) = \mathbb{E}_{\tau \sim \pi_{\theta}}\left[ (R(\tau) - b) \sum_{t=0}^{T-1} \nabla_{\theta} \log \pi_{\theta}(\mathbf{a}_t \mid \mathbf{s}_t)\right],
\end{equation}
where $b \in \mathbb{R}$ is a constant, and subtracting $b$ can reduce the variance in practice.

In the FinRL tasks, the trading strategy is governed by $\pi_{\theta}$ to maximize $J(\theta)$. When dealing with complex and noisy financial datasets, achieving a low-variance gradient estimation $\nabla J(\theta)$ is crucial for stable and reliable policy updates, reducing the risk of suboptimal trading strategies.

To estimate $\nabla J(\theta)$ in (\ref{eq:gradient_J}), we can use the Monte Carlo method \cite{monte_carlo}
\begin{equation}
    \label{eq:gradient_esitmate}
    \nabla J(\theta) = \frac{1}{n} \sum_{j=1}^{n} \left( (R(\tau^{(j)}) - b)  \sum_{t=0}^{T-1}  ~\nabla_{\theta} \log \pi_{\theta}(\mathbf{a}_t^{(j)} \mid \mathbf{s}_t^{(j)}) \right), 
\end{equation}
where $n$ trajectories are used. The Law of Large Numbers guarantees that as the sample size $n$ increases, the estimation of $ \nabla J(\theta)$ will converge to its expected value. According to the Central Limit Theorem, increasing $n$ leads to a reduction in the variance of the estimate.

As shown in (\ref{eq:gradient_esitmate}), a large number $n$ of trajectories sampled during the simulation phase is required to reduce the variance of $\nabla J(\theta)$. 
%There exists a high degree of parallelism in $n$. 
In (\ref{eq:gradient_esitmate}), each $j$ in the outer sum from $1$ to $n$ corresponds to a separate trajectory $\tau^{(j)}$, which can be considered as a complete and independent simulation of the policy $\pi_\theta$ in the environment. Therefore, each trajectory $\tau^{(j)}$ can be simulated in parallel, allowing for a high degree of parallelism. In FinRL, parallel simulation involves executing the trading strategy in multiple market scenarios simultaneously. The parallelism accelerates the simulation phase, allowing for more rapid updates and iterations of the policy $\pi_{\theta}$. Therefore, the trading strategy governed by $\pi_{\theta}$ can be quickly updated and adapted to changing market conditions.

% \end{document}

% \subsection{MDP Formulation}
% MDP

% \subsubsection{Parallelism}

% \subsection{Challenges}
% Policy instability; sampling bottleneck; signals from financial documents.

% \subsection{FinRL Contest 2023-2025}
% Design principle. 

% Typical Tasks (stock trading, ensemble learning for cryptocurrency trading, LLM-generated signals in FinRL).

%\section{Market Data and Environments}
\section{FinRL Environments}

% \documentclass[../main.tex]{subfiles}
% \graphicspath{{\subfix{../fig/}}}
% \begin{document}

Market data and the simulation environment are two critical components of FinRL that influence the definition of tasks. In this section, we provide detailed information on market data and GPU-optimized parallel market environments.

\subsection{Automated Data Curation Pipeline for Dynamic Market Data}
\label{sec:paper_trading}

We first describe different sources of dynamic market data used in FinRL Contests, along with feature engineering, market data split strategies, and the automated data curation pipeline.

\begin{table}[ht]
\centering
\begin{tabular}{|l|l|p{9cm}|}
\hline
\textbf{Indicator} & \textbf{Name} & \textbf{Description} \\
\hline
macd & Moving Average Convergence Divergence & A trend-following momentum indicator that shows the relationship between two exponential moving averages (EMAs) of a stock’s price. Traders use the MACD to identify potential trend changes, divergence between price and momentum, and overbought or oversold conditions. \\
\hline
boll\_ub & Bollinger Bands Upper Band & Bollinger Bands are used to visualize the volatility and potential price levels of a stock. The upper band represents the upper volatility boundary, showing where the price might find resistance. \\
\hline
boll\_lb & Bollinger Bands Lower Band & Similarly, the lower band represents the lower volatility boundary and shows where the price might find support. \\
\hline
rsi\_30 & Relative Strength Index for 30 periods & A momentum oscillator that measures the speed and change of price movements. RSI oscillates between zero and 100. \\
\hline
cci\_30 & Commodity Channel Index for 30 periods & A versatile indicator that can be used to identify a new trend or warn of extreme conditions. It measures the current price level relative to an average price level over a given period of time. \\
\hline
dx\_30 & Directional Movement Index for 30 periods & An indicator that assesses the strength and direction of a trend of a stock. It does this by comparing highs and lows over time. \\
\hline
close\_30 & 30-Period Simple Moving Average of Closing Prices & Represents the average closing price over the last 30 periods. This moving average provides a smoothed representation of the asset's price over the 30 periods, making it easier to identify trends and potential support/resistance levels. \\
\hline
close\_60 & 60-Period Simple Moving Average of Closing Prices & Represents the average closing price over the last 60 periods. This moving average provides a smoothed representation of the asset's price over the 60 periods, making it easier to identify trends and potential support/resistance levels. \\
\hline
vix & Volatility Index & Often referred to as the "fear index", it represents the market's expectation of 30-day forward-looking volatility. It is calculated from the prices of selected stock option contracts on the S\&P 500 Index. \\
\hline
turbulence & Turbulence & To control the risk in a worst-case scenario, such as the financial crisis of 2007–2008, FinRL employs the financial turbulence index that measures extreme asset price fluctuation. \\
\hline
\end{tabular}
\caption{Market indicators and their descriptions}
\label{tab:technical_indicators}
\end{table}

\textbf{Data Sources.} We leverage both structured market data and unstructured financial documents. Structured market data captures quantitative market behavior, such as historical price information and market indicator data. Unstructured financial documents provide information about economic trends, regulations, and events that drive market movements. The structured market data is processed into price features $\mathbf{p}_t$, which are included in state $\mathbf{s}_t$. We outline the financial data utilized in FinRL as follows:
\begin{itemize}
    \item \textbf{OHLCV data}. OHLCV (open, high, low, close, volume) data is typical historical volume-price data in finance. We provide daily OHLCV data for stocks from 1999 to 2023. This dataset is prepared through \texttt{yfinance}\footnote{\url{https://yfinance-python.org/}} (an open-source Python library) and FinRL-Meta \cite{liu2022finrl, liu2024dynamic}, released under the CDLA-permissive-2.0 license. The missing values are removed. We also provide data APIs to download market data, such as Alpaca\footnote{\url{https://alpaca.markets/}}, which are held on the open-source repository FinRL-Meta\footnote{\url{https://github.com/AI4Finance-Foundation/FinRL-Meta}} \cite{liu2022finrl, liu2024dynamic}.
    \item \textbf{Limit order book (LOB)} data at second level for cryptocurrency trading\footnote{\url{https://www.kaggle.com/datasets/martinsn/high-frequency-crypto-limit-order-book-data/data}}. LOB data offers a detailed view of market depth and liquidity, capturing the behavior of market participants and providing valuable insights into market trends. Given its high granularity and extensive size, this data is ideally suited for participants to leverage massively parallel simulations, enabling more effective development of cryptocurrency trading strategies. %In addition, we also provide 8 strong factors extracted from adapted 101 formulaic alphas \cite{alpha101} through a recurrent neural network (RNN)\footnote{\url{https://github.com/Open-Finance-Lab/FinRL_Contest_2025/tree/main/Task_2_FinRL_AlphaSeek_Crypto}}.
    \item \textbf{Financial news}. The Financial News and Stock Price Integration Dataset (FNSPID) \cite{dong2024fnspid} contains 15 million time-aligned financial news articles from 1999 to 2023 for constituent companies in S\&P 500. It is released under CC-BY-NA 4.0 for academic purposes. From the FNSPID dataset, we randomly select one news article per day per stock. The news is aggregated with OHLCV data to ensure temporal alignment and create a multimodal dataset. For cryptocurrency news, we aggregate BTC news collected from multiple sources \footnote{The aggregated BTC news dataset: \url{https://huggingface.co/datasets/SecureFinAI-Lab/FinRL_BTC_news_signals}. Data sources: \url{https://huggingface.co/datasets/edaschau/bitcoin_news}; \url{https://github.com/soheilrahsaz/cryptoNewsDataset}} and ensure the temporal alignment with the LOB data. We also provide data APIs to access financial news, such as Yahoo Finance and Finnhub. These data APIs are held on the open-source repository FinGPT\footnote{\url{https://github.com/AI4Finance-Foundation/FinGPT}} \cite{liu2023fingpt}.
\end{itemize}

% This multimodal combination requires careful time alignment.
% \textbf{Feature Engineering.} Before the dataset can be used in contests, we perform preprocessing and feature enhancement. For the OHLCV dataset, the missing values are removed. Based on daily price data, ten market indicators listed in Table \ref{tab:technical_indicators} are computed to enrich the feature vector $\mathbf{f}_t$. For the LOB data, bid and ask prices at different market depths are extracted, and 101 adapted alpha factors \cite{alpha101} are constructed. An RNN is trained to distill 8 strong factors from the 101 weak alphas, modeling and predicting the prices. For the FNSPID dataset, we randomly select one news article per day per stock. The news is aggregated with OHLCV data to ensure temporal alignment.

\textbf{Feature Engineering.} We conduct feature engineering before making the dataset available for contests.
\begin{itemize}
    \item \textbf{Market indicators}. Ten market indicators listed in Table \ref{tab:technical_indicators} are computed based on OHLCV data. These indicators are used to enrich the feature vector $\mathbf{f}_t$ in state $\mathbf{s}_t$.
    \item \textbf{ML-learned factors}. For the LOB data, bid and ask prices at different market depths are extracted, and 101 adapted alpha factors \cite{alpha101} are constructed. An RNN is trained to distill 8 strong factors from the 101 weak alphas, modeling and predicting the prices. These factors are used to enrich the feature vector $\mathbf{f}_t$ in state $\mathbf{s}_t$.
\end{itemize}
The feature selection process involves two steps: 1) compute the Pearson correlation matrix of the features, and 2) select one representative feature from each group of highly correlated features, following the method proposed by Gort et al \cite{gort2022deep}.

\textbf{LLM-Engineered Signals}. LLMs have been envisioned to be very promising for generating alpha signals from multiple financial data sources \cite{nie2024llmsurevey, guo2024quant, cao2025deep}. We encourage participants to leverage LLMs and train a trading agent for making informed decisions. Here we show the sentiment and risk scores generated by DeepSeek models from financial news \cite{deepseek2025mostapha}:

% Don't use deepseek at the beginning. 
\begin{itemize}
    \item \textbf{Sentiment score}. An LLM assigns a sentiment score $u$ of 1 to 5 according to the news, with 1 for strongly negative and 5 for highly positive. For example, we can use DeepSeek-V3 \cite{liu2024deepseek} to generate signals. It is included in the feature vector $\mathbf{f}_t$ and is used to adjust actions via the sentiment factor $l^i_t = 1 + 0.05(u-3) \text{sign}(a^i_t)$. The factor is close to 1 for the stability of the algorithm. The adjusted action is $a_t^{i'} = l^i_ta_t^i$, which is amplified under positive sentiment and dampened under negative sentiment.
    \item \textbf{Risk level}. An LLM assigns a risk level $q$ of 1 to 5 from the news, with 1 for low risk and 5 for high risk. It is included in the feature vector $\mathbf{f}_t$ and is used to penalize rewards through the risk factor $m_t^i = 1+0.05(q_t^i-3)$. The aggregated risk factor is $M_t = \sum_i^{K}w_t^i{m}_t^i$, where $w_t^i$ is the portfolio weight of the stock $i$ and $\sum w^i = 1$. $M_t > 1$ penalizes the reward for high risk; $M_t < 1$ increases the reward for low risk.
\end{itemize}

% Training, validation, and testing data. (How to prevent future data leakage)

\textbf{Automated Data Curation Pipeline.} After preparing the datasets, we split them into two parts, one released to participants for training and the other held by organizers for evaluation purposes. We use the temporal partitioning approach to maintain the temporal integrity of financial data and prevent future information leakage.
\begin{itemize}
    \item \textbf{Dataset released to participants}. We provide long-span historical datasets for participants to develop their models. It covers different market patterns, such as financial crises, bull markets, and bear markets. 
    %Different strategies can be used to split the data, such as static splits (70\% training, 30\% validation) and rolling windows with periodic retraining. 
    Participants are not restricted to the provided datasets and APIs. They are encouraged to use external data sources, such as alternative market indicators, financial news feeds, and tweets.
    \item \textbf{Dataset withheld for evaluation}. Depending on the task and data availability, we use two methods to obtain the evaluation dataset: 1) we extract the most recent $\sim$15\% of the original dataset, withhold it as the evaluation dataset, and encrypt all timestamps; 2) we collect out-of-sample data for the period after the model submission deadline. It involves downloading market data and financial news via \texttt{yfinance} or other APIs. These two methods will avoid future data leakage and ensure a fair assessment. This setup ensures that the evaluation simulates real-world, time-forward deployment, where models should generalize to new, unseen market data.
\end{itemize}

% \textbf{For training and validation}, we provide historical datasets. Participants are not restricted to the provided datasets and APIs. They are encouraged to use external data sources, such as alternative market indicators, financial news feeds, and tweets. \textbf{For evaluation}, we use \texttt{yfinance} to download financial news and OHLCV data after the model submission deadline. This will avoid future data leakage and ensure a fair assessment. This setup ensures that the evaluation simulates real-world deployment, where models should generalize to new, unseen market data.

To better reflect real-world forward-moving financial markets, we present an automated data curation pipeline. We illustrate this pipeline using the example of the paper trading task. The conventional backtesting approach splits historical market data into in-sample and out-of-sample time periods. Backtesting may be performed multiple times on the out-of-sample data. However, stock paper trading is strictly required to be carried out \textbf{once} in real-time on the real-world market.

\begin{figure}[t]
\centering
\includegraphics[width=0.7\linewidth]{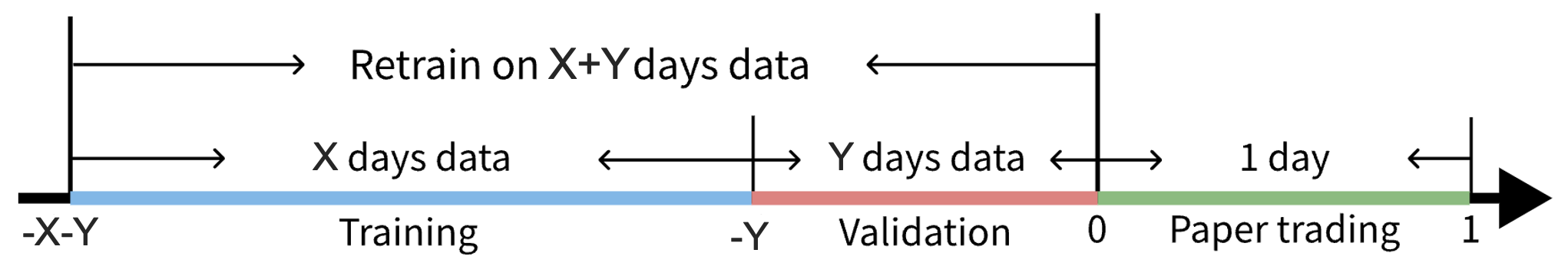}
\caption{Data split for a window with training, validation, and trading.}
\label{fig:rw_split}
\end{figure}

%%%%%%%%%%%%%%%%%%%%%%%%%%%%%%%%%%%%%%%%%%%%%%%%%%%%%%%%%%%%%%

\begin{algorithm}[t]
	\caption{Pseudo code for stock paper trading}
	\label{algo_pt}
	\begin{algorithmic}[1]
	    \State Initialize a set of hyperparameters;
	    \For {$z = 0$ to $Z-1$}
	    \State \# \textbf{Step 1)}. Train an agent
		\State Train agent on data period $[z{-}Y{-}X,\ z{-}Y{-}1]$
        \State Validate trained agent and tune hyperparameters on data period $[z{-}Y,\ z{-}1]$
        \State \# \textbf{Step 2)}. Retrain the agent
        \State Retrain agent on data period $[z{-}Y{-}X,\ z{-}1]$ using tuned hyperparameters
        \State \# \textbf{Step 3)}. Perform paper trading for one day
        \State Trade on day $z$ with trained agent
        \EndFor
	\end{algorithmic}
\end{algorithm}

Fig. \ref{fig:rw_split} shows our ``training-validation-trading" pipeline. In a window, there are $X$ days' data for training and $Y$ days' data for validation. At the end of a window, we perform paper trading for $1$ day. Note that we always retrain the agent using $X+Y$ days of training and validation data together. Then, we roll the window forward by $1$ day ahead and perform the above steps for a new window. A paper trading is always carried out for $1$ day. Therefore, $Z$ windows correspond to $Z$ trading days.

Alg. \ref{algo_pt} summarizes the pipeline of paper trading. For $Z$ trading days ($z = 0, 1, ..., Z-1$), we keep doing the following three steps:
\begin{itemize}
    \item \textbf{Step 1)}. Download and process $X$-day data, from day $z-Y-X$ to day $z-Y-1$. Then build the data into a gym-style environment and train the agent. Then download and process Y-day data, from day $z-Y$ to day $z-1$. Then build the data into a gym-style environment and validate how the agent performs. According to the agent's performance on the validation environment, adjust the hyperparameters.
    \item \textbf{Step 2)}. Build the training and validation data, totally $X+Y$ days from day $z-Y-X$ to day $z-1$,
    %(note that there are $390$ data points for each day's minute level data)
    into a gym-style environment. Update hyperparameters to the values chosen from \textbf{Step 1)}. Then retrain the agent on this $X+Y$-day environment.
    \item \textbf{Step 3)}. Deploy the trained agent to the paper trading market. %and trade from 9:30am to 4:00pm.
\end{itemize}
Before each trading day $z$, we use the historical data period from $z-Y-X$ to $z-Y-1$ to train the FinRL agent. Then, data periods from $z-Y$ to $z-1$ are used to validate the trained agent, adjusting hyperparameters accordingly. We retrain the agent by combining all the data from the training and validation period, that is, the data period from $z-Y-X$ to $z-1$. Finally, we perform $1$ day paper trading on the $z$-th day. We continue this process for $z = 0, 1, ..., Z-1$.

\subsection{Market Environments}
The financial data is processed into the standard gym-style market environments to build a standard practical environment. This part describes the market environments with near-real market constraints. To address the sampling bottleneck of the training stage, we develop massively parallel market environments on GPUs.

\subsubsection{Market Environment with Trading Constraints}
% add details about these functions.

After processing the financial data into standard gym-style market environments, we define the key operations in the FinRL market environment, including: 
\begin{itemize}
    \item {  \texttt{reset}}: $\mathbf{s}_t \rightarrow \mathbf{s}_0$, resets the environment to its initial state.
    \item {  \texttt{step}}: ($\mathbf{s}_t, \mathbf{a}_t) \rightarrow \mathbf{s}_{t+1}$, executes $\mathbf{a}_t$ and updates $\mathbf{s}_t$ to $\mathbf{s}_{t+1}$.
    \item {  \texttt{reward}}: $(\mathbf{s}_t, \mathbf{a}_t, \mathbf{s}_{t+1}) \rightarrow r_{t+1}$, computes the reward.
\end{itemize}

\noindent To reflect the real-world trading scenario, we incorporate near-real trading constraints:
\begin{itemize}
    \item \textbf{Transaction costs}. We set a cost of $0.1\%$ for each action \{buy, sell\}, accounting for commission and slippage.
    \item \textbf{Market volatility}. The turbulence index and VIX index are risk indicators. A large value signals heightened volatility from factors like investor fear and increased uncertainty, while a small value signals increased stability in markets. 
\end{itemize}

\subsubsection{Massively Parallel Simulations on GPUs}\label{sec:massively}

\begin{figure}[t]
    \centering
    \includegraphics[width=0.35\linewidth]{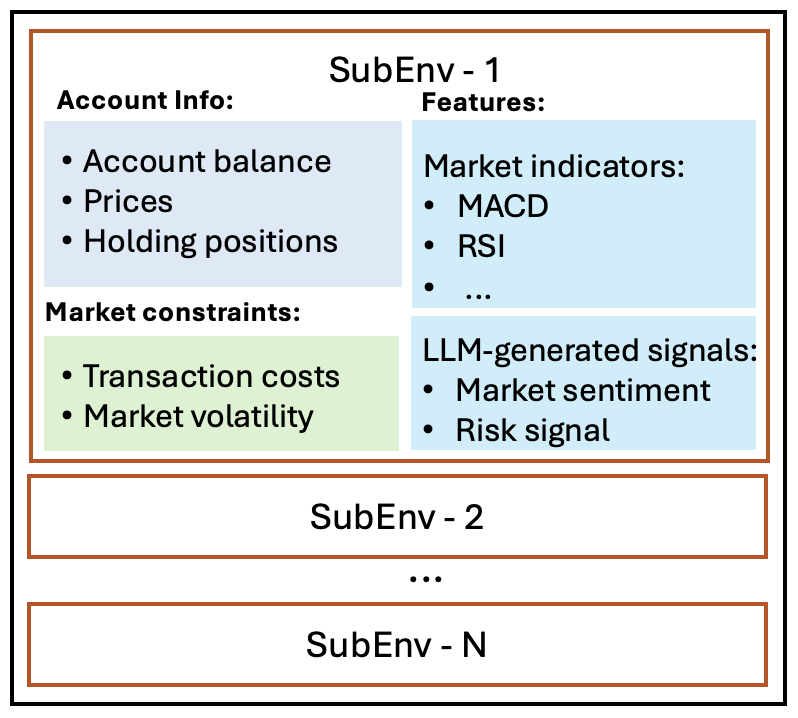}
    \caption{Vectorized environment.}
    \label{fig:vec_env}
\end{figure}

% (figures and references)
Stable policy updates require a low-variance gradient estimate $\nabla J(\theta)$ from a large $n$ independent trading trajectories $\tau$, where there is high parallelism, as shown in (\ref{eq:gradient_esitmate}). PyTorch's \texttt{vmap} can map operations over some dimension onto parallel GPU cores, exploiting the parallelism of (\ref{eq:gradient_J}). Therefore, we construct vectorized environments and provide GPU optimization via \texttt{vmap}. As shown in Fig. \ref{fig:vec_env}, a vectorized environment manages parallel sub-environments (SubEnvs). 

Using \texttt{vmap}, the \texttt{step} and \texttt{reward} functions are vectorized to operate in massively parallel environments. For example, the \texttt{reward} function, vectorized by \texttt{vmap}, computes the reward on $(\mathbf{s}_t, \mathbf{a}_t, \mathbf{s}_{t+1})$ for each SubEnv simultaneously. This computation is dispatched to available GPU cores, each responsible for calculating its assigned data.

Data samples are stored as tensors in GPU memory. They have the shape $N \times T \times D$:
\begin{itemize}
    \item $N$ is the number of parallel SubEnvs.
    \item $T$ is the number of steps in a trajectory.
    \item $D$ is the dimension as in Section \ref{sec:mdp}, where $D = K(I+2)+1$ for state, $D = K$ for action, and $D = 1$ for reward.
\end{itemize}
The tensors for states ($\mathbf{s} \in \mathbb{R}^{K(I+2)+1}$), actions ($ \mathbf{a} \in \mathbb{R}^{K}$), and rewards ($r \in \mathbb{R}$) are as follows:
\[
\begin{bmatrix}
    %\mathbf{s}_0^1 & \mathbf{s}_0^2 & \cdots & \mathbf{s}_0^N \\
    \mathbf{s}_0^1 &  \mathbf{s}_1^1 &  \cdots &  \mathbf{s}_{T-1}^1 \\
    \mathbf{s}_0^2 &  \mathbf{s}_1^2 &  \cdots &  \mathbf{s}_{T-1}^2 \\
    \vdots &  \vdots &  \ddots &  \vdots \\
    \mathbf{s}_0^N &  \mathbf{s}_1^N &  \cdots &  \mathbf{s}_{T-1}^N \\
\end{bmatrix},
\begin{bmatrix}
    %\mathbf{a}_0^1 & \mathbf{a}_0^2 & \cdots & \mathbf{a}_0^N \\
    \mathbf{a}_0^1 &  \mathbf{a}_1^1 &  \cdots &  \mathbf{a}_{T-1}^1 \\
    \mathbf{a}_0^2 &  \mathbf{a}_1^2 &  \cdots &  \mathbf{a}_{T-1}^2 \\
    \vdots &  \vdots &  \ddots &  \vdots \\
    \mathbf{a}_0^N &  \mathbf{a}_1^N &  \cdots &  \mathbf{a}_{T-1}^N \\
\end{bmatrix},
\begin{bmatrix}
    %r_0^1 & r_0^2 & \cdots & r_0^N \\
    r_1^1 &   r_2^1 &  \cdots &  r_{T}^1 \\
    r_1^2 &   r_2^2 &  \cdots &  r_{T}^2 \\
    \vdots &  \vdots &  \ddots &  \vdots \\
    r_1^N &  r_2^N &  \cdots &  r_{T}^N \\
\end{bmatrix}.
\]
% \[
% \begin{bmatrix}
%     %\mathbf{s}_0^1 & \mathbf{s}_0^2 & \cdots & \mathbf{s}_0^N \\
%     \mathbf{s}_1^1 &  \mathbf{s}_2^1 &  \cdots &  \mathbf{s}_T^1 \\
%     \mathbf{s}_1^2 &  \mathbf{s}_2^2 &  \cdots &  \mathbf{s}_T^2 \\
%     \vdots &  \vdots &  \ddots &  \vdots \\
%     \mathbf{s}_1^N &  \mathbf{s}_2^N &  \cdots &  \mathbf{s}_T^N \\
% \end{bmatrix},
% \begin{bmatrix}
%     %\mathbf{a}_0^1 & \mathbf{a}_0^2 & \cdots & \mathbf{a}_0^N \\
%     \mathbf{a}_1^1 &  \mathbf{a}_2^1 &  \cdots &  \mathbf{a}_T^1 \\
%     \mathbf{a}_1^2 &  \mathbf{a}_2^2 &  \cdots &  \mathbf{a}_T^2 \\
%     \vdots &  \vdots &  \ddots &  \vdots \\
%     \mathbf{a}_1^N &  \mathbf{a}_2^N &  \cdots &  \mathbf{a}_T^N \\
% \end{bmatrix},
% \begin{bmatrix}
%     %r_0^1 & r_0^2 & \cdots & r_0^N \\
%     r_1^1 &   r_2^1 &  \cdots &  r_T^1 \\
%     r_1^2 &   r_2^2 &  \cdots &  r_T^2 \\
%     \vdots &  \vdots &  \ddots &  \vdots \\
%     r_1^N &  r_2^N &  \cdots &  r_T^N \\
% \end{bmatrix}.
% \]
Storing data samples as tensors in GPU memory bypasses the CPU-GPU bandwidth bottleneck.

\begin{figure}[h]
    \centering
    \includegraphics[width=0.6\linewidth]{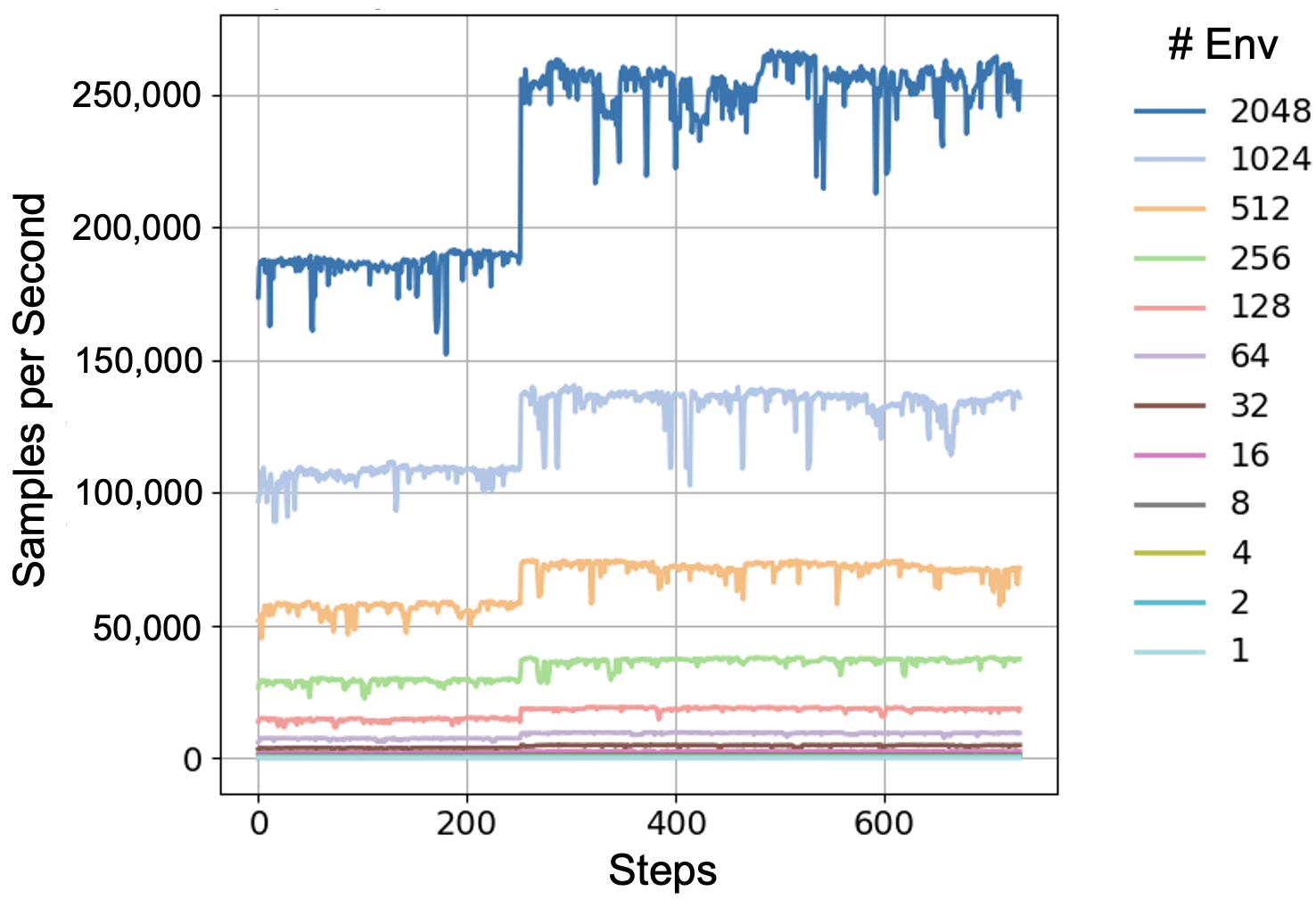}
    \caption{Samples per second for the stock trading task (using NVIDIA A100 GPU).}
    \label{fig:samples_stock}
\end{figure}

\textbf{Improved Sampling Speed with Massively Parallel Environments on GPU}. We evaluated the sampling speed measured in samples per second using vectorized environments for stock trading. We used the PPO agent and the OHLCV data of $30$ constituent stocks in the Dow Jones index, from 2020-01-01 to 2022-01-01. The NVIDIA A100 GPU is used. The numbers of parallel environments vary from $1$, $2$, $4$, $\ldots$, to $2,048$. As shown in Fig. \ref{fig:samples_stock}, the average sampling speed with $2,048$ parallel environments is $227,212.54$ samples per second. The sampling speed is improved by $1,649.93\times$ compared with a single environment. The sampling speed scales approximately linearly with the number of parallel environments. The results show the effectiveness of massively parallel simulation in improving sampling speed in FinRL Contest tasks.

% \end{document}

% \subsection{Dynamic Market Data}
% OHLCV, LOB, financial news.

% Training, validation, and testing data. (How to prevent future data leakage)

% \subsubsection{Automated Data Curation Pipeline}

% paper trading

% \subsection{Market Environments}

% \subsubsection{Market Constraints}

% \subsubsection{LLM-Generated Signals}

% \subsubsection{Massively Parallel Simulations on GPUs}

% (figures and references)

% \subsection{Baseline Methods}

% \subsubsection{Market Index and Mean-Variance  Method}

% \subsubsection{Ensemble Methods}

\section{Benchmarking Protocol}
To assist participants in becoming familiar with FinRL contests and to ensure fair comparisons, we establish a benchmarking protocol and provide several baseline methods.

% \documentclass[../main.tex]{subfiles}
% \graphicspath{{\subfix{../fig/}}}
% \begin{document}

\subsection{Baseline Methods}

\subsubsection{Market Index and Mean-Variance Method}

We use market indexes and the mean-variance optimization strategy as baselines. They are widely adopted in the financial industry.
\begin{itemize}
    \item \textbf{DJIA}. The Dow Jones Industrial Average index is a price-weighted index for 30 blue-chip U.S. companies. It is calculated as the sum of constituent stock prices divided by the Dow Divisor. The divisor is a constant adjusted for stock splits and structural changes. The data can be downloaded by \texttt{yfinance}. DJIA is one of the oldest and most recognized market indexes. It provides a real-world baseline to evaluate whether a FinRL agent outperforms a passive investment strategy.
    \item \textbf{S\&P 500}. The Standard and Poor's 500 is a capitalization-weighted index tracking the stock performance of 500 leading companies. The data can be downloaded by \texttt{yfinance}. Covering approximately 80\% of the total market capitalization, it offers a broader and more diversified baseline.
    \item \textbf{Mean-variance optimization}. Mean-variance optimization strategy, as part of Modern Portfolio Theory (MPT) \cite{mdp1952markowitz}, constructs portfolios that maximize the expected return for a given level of risk. It uses expected asset returns and covariances to solve an optimization problem. We typically use the past one year's daily price data to calculate expected returns and the covariance matrix. We limit individual stock weights to a maximum of 5\%. Mean-variance optimization is a foundational technique in portfolio management and can serve as a classical financial optimization strategy baseline.
\end{itemize}

\subsubsection{Ensemble Agent}

Ensemble methods have shown effectiveness in enhancing overall performance by combining selected actions or action probabilities from component RL algorithms ~\cite{wiering2008ensemble}. We train ensemble trading agents to mitigate policy instability.

\paragraph{Agent Diversity} \label{sec:agent_diversity}
%As shown in Fig. \ref{fig:ensemble_method}, for different tasks, we have various types of agents, each with multiple instances, to ensure a large enough number of agents. 
%We use various approaches to ensure agent diversity, which is crucial to enhancing the ensemble's performance.
The diversity of component agents is essential for risk mitigation by leveraging various trading agents. Achieving high diversity requires training multiple agents across environments that simulate different market scenarios.

\textbf{Using KL divergence in objective functions}. To enforce diversity among the component agents, we introduce a Kullback-Leibler (KL) divergence term into the agent's objective function. The KL divergence measures how one probability distribution diverges from another ~\cite{perez2008kullback}. The KL divergence term penalizes similarities in policies between different agents, encouraging them to converge on different trading strategies. The new objective function for a component agent is as follows:
\begin{equation}
    \max ~~L_{\text{new}}(\theta_A)=L_{\text{original}}(\theta_A) + \lambda \sum_{A \neq B} \text{KL}(\pi_{\theta_B}||\pi_{\theta_A}),
\end{equation}
where $\theta_A$ are the policy parameters for agent $A$, $L(\theta_A)$ is the objective function, $\text{KL}(\pi_{\theta_B}||\pi_{\theta_A})$ is the KL divergence between agent $A$'s and agent $B$'s policies, and $\lambda > 0$ is a regularization constant. After obtaining the trained agents, one can ensemble them using a majority voting method or the weighted sum method \cite{yang2020deep}. 

\textbf{Using various datasets}. The financial datasets used for training component agents are varied. For each stock or crypto, a random percentage change ranging from $-1\%$ to $1\%$ is generated and applied to its prices, which shifts the price scale while preserving the original price trends.
%It helps prevent models from overfitting. 
%We also use different periods of data to train the component agents. 
Agents are also trained on different stocks from the test set for the stock trading task. It enables agents to learn various strategies for a broader range of stocks rather than reacting to a limited number of stocks.

% \subsubsubsection{Ensemble Methods for FinRL Tasks}
% % As shown in Fig. \ref{fig:ensemble_method}, we use different agents and ensemble methods for stock and crypto trading tasks.

\begin{figure*}[t]
  \centering
  \includegraphics[width=\textwidth]{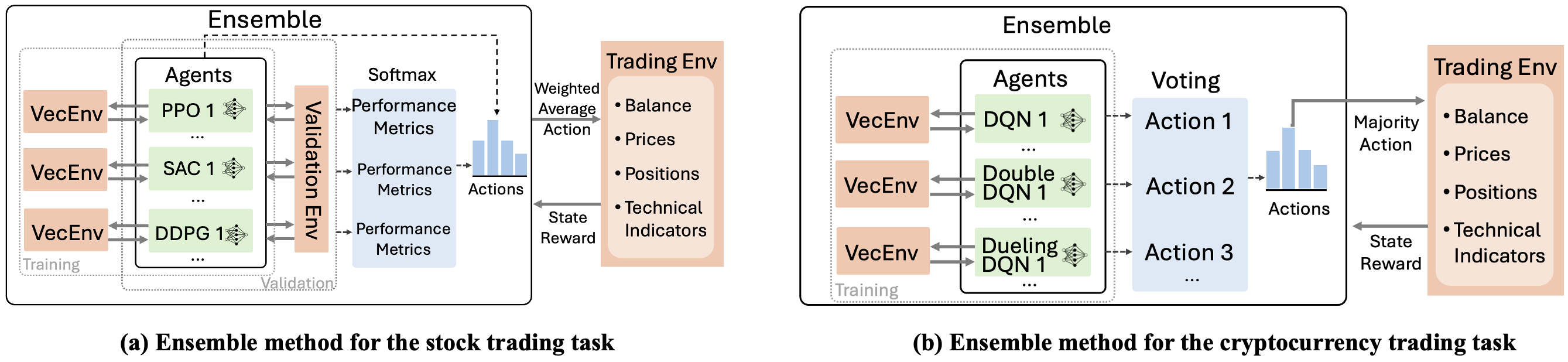}
  \caption{Ensemble methods.}
  \label{fig:ensemble_method}
\end{figure*}

\paragraph{\textbf{Stock Trading Task}} \label{sec:stock_ensemble}
As shown in Fig. \ref{fig:ensemble_method} (a), the ensemble includes PPO, SAC, and DDPG agents. The ensemble's final trading action is determined by weighted averaging over the agents' action probabilities. The process is as follows:
\begin{itemize}
    \item \textbf{Traning}. Agents are trained independently with a VecEnv on a 30-day training rolling window, using massively parallel simulation in Section \ref{sec:massively} and agent diversity methods in Section \ref{sec:agent_diversity}. 
    % We use different strategies to ensure agent diversity as described in \ref{sec:agent_diversity}.
    \item \textbf{Validation}. After training, agents are validated on a 5-day rolling window. Sharpe ratios are calculated to evaluate their ability to balance returns with associated risks.
    \begin{equation} \label{sharpe_ratio}
        \text{Sharpe Ratio} = \frac{\bar{r}_p - r_f}{\sigma_p},
    \end{equation}
    where $\bar{r}_p$ is the portfolio return, $r_f$ is a chosen risk-free rate, and $\sigma_p$ is the standard deviation of the portfolio return.
    \item \textbf{Weights calculation}. Agents with very low Sharpe ratios are discarded. Weights for the remaining agents are calculated using a softmax function applied to their Sharpe ratios.
    \item \textbf{Trading}. The ensemble acts based on a weighted average of agent action probabilities during a 5-day trading window.
\end{itemize}
%After each loop, the training, validation, and trading windows are rolled forward. 
This rolling window approach ensures that the ensemble method remains adaptive to the continuously changing market. 

%The stock trading task is designed as a portfolio trading task, where the model predicts the number of shares to buy, sell, and hold for any stock at each timestep. As seen in Fig. \ref{fig:ensemble_method} we use PPO, SAC ~\cite{ haarnoja2018sac}, and DDPG \cite{ddpg} agents. These models were trained independently on the in-sample data to develop distinct trading strategies using a rolling window where the ensemble agents are trained on $30$ days of data, then evaluated on $5$ out of sample days and then tested on $5$ more out of sample days on a repeating basis. Post-training, agents were validated based on their ability to balance returns with associated risks. Weights for each agent’s outputs in the ensemble were assigned using a softmax function applied to the Sharpe ratios obtained during this validation phase. 

\paragraph{\textbf{Crypto Trading Task}} \label{sec:crypto_ensemble}
For crypto trading at a relatively high frequency, market movements can be modeled as discrete events, which require a discrete action space. As shown in Fig. \ref{fig:ensemble_method} (b), DQN ~\cite{Mnih2015}, Double DQN ~\cite{doubledqn}, and Dueling DQN ~\cite{duelingdqn} are used to handle this discrete action space. In addition, the dataset for a single crypto is relatively small. DQN and its variants, with fewer parameters and simpler architectures, can be trained faster to avoid overfitting. Moreover, trading at a high frequency requires fast responses, and DQN agents can offer lower latency in decision-making compared to more complex models. The ensemble model uses majority voting to combine the actions of component agents. Majority voting ensures the chosen action reflects consensus among agents, mitigating biases from any single agent's actions ~\cite{ensemble_ganaie}. The process is as follows:
\begin{itemize}
    \item \textbf{Training}. Each component agent is independently trained with a VecEnv, using the massively parallel simulation in Section \ref{sec:massively} and the agent diversity methods in Section \ref{sec:agent_diversity}. 
    \item \textbf{Action ensemble and trading}. During the trading phase, each agent processes the same market state and determines an action based on its policy. The majority action is selected as the final ensemble action.
\end{itemize}

\subsection{Evaluation Protocol}

% How to ensure fair evaluation (same platform, same setting, evaluators
\begin{figure}[t]
\centering
\includegraphics[width=0.8\linewidth]{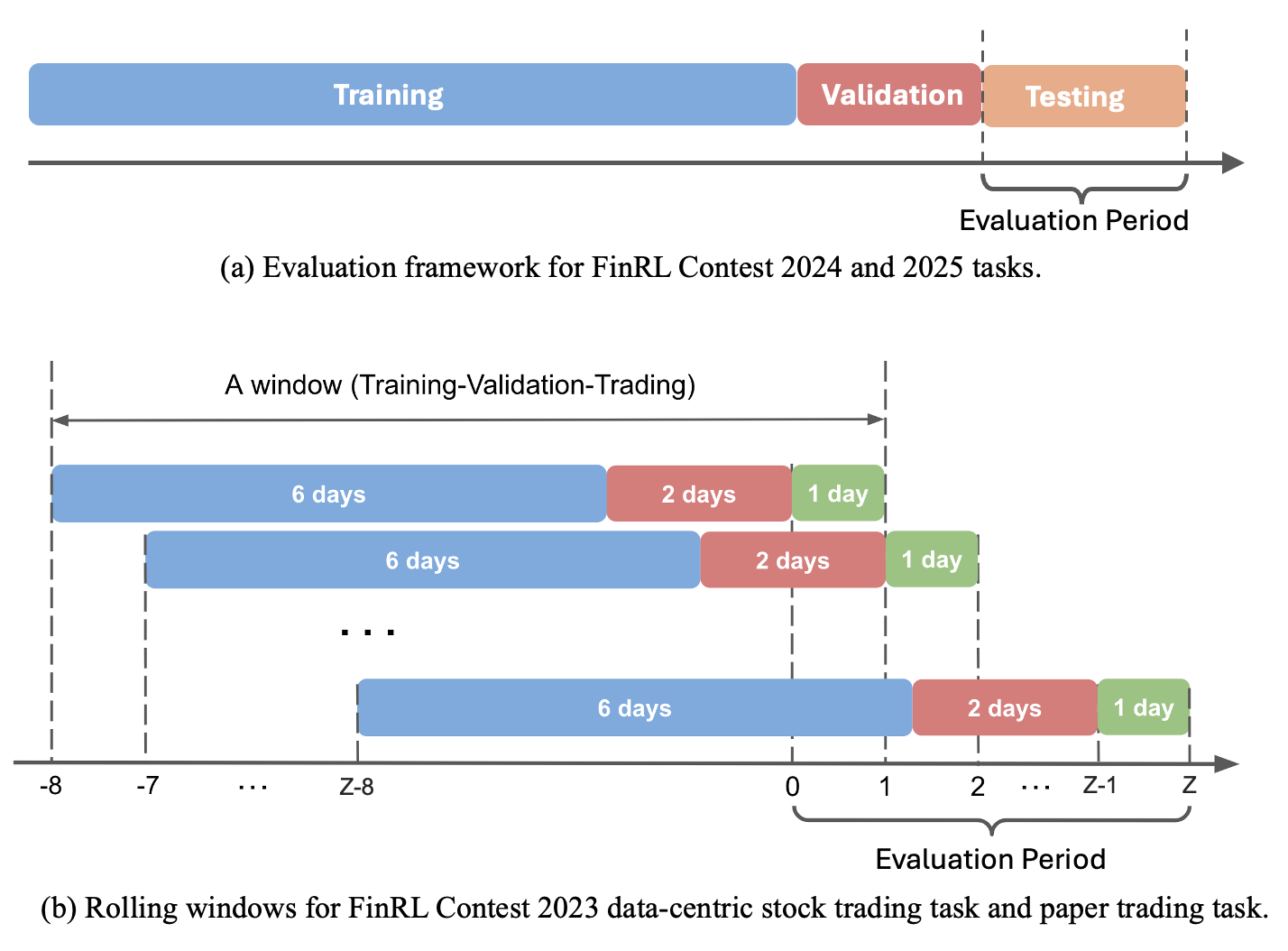}
\caption{The rolling windows for the evaluation of the stock trading task.}
\label{fig:full_process}
\end{figure}

To evaluate the models, we adopt two frameworks, backtesting and rolling windows, as shown in Fig. \ref{fig:full_process}:
\begin{itemize}
    \item \textbf{Backtesting}. We use backtesting to evaluate models for trading tasks in FinRL Contests 2024 and 2025, as shown in Fig. \ref{fig:full_process} (a). The datasets released to participants are used for their training and validation. The withheld out-of-sample dataset is used for evaluation. We did not use the rolling window framework because the evaluation datasets for stock trading have long time spans, and the crypto data is high-frequency. In both cases, retraining models frequently would be too time-consuming and difficult to manage.  
    \item \textbf{Rolling windows}. We use the rolling window framework to evaluate models for FinRL Contest 2023 data-centric stock trading and paper trading, as discussed in Section \ref{sec:paper_trading}. Fig. \ref{fig:full_process} (b) presents the mechanism of each rolling window for the stock trading task. There are in total $9$ days in a rolling window. We divide it into three stages, namely, $6$ days of training, $2$ days of validation, and $1$ day of trading using the trained DRL agent. The models are retrained for $8$ days before each trading day. The validation stage allows adjusting the hyperparameters or selecting an agent from the ensemble. Finally, the well-trained agent performs stock trading on the last day of a rolling window.
\end{itemize}

To ensure fair and reproducible benchmarking for all models, the evaluation process follows the principles below:
\begin{itemize}
    \item \textbf{Uniform evaluation platform}. All models are evaluated on the same infrastructure to eliminate differences caused by hardware or software variations. Specifically, we conduct evaluations using Google Cloud Platform (GCP) virtual machines with identical hardware configurations. The software environment is Ubuntu 22.04.2 LTS and Python 3.10.12.
    \item \textbf{Controlled evaluation setting}. All models are evaluated in the same environment settings, including testing data, initial investment amount, and transaction costs. No additional training or model adjustment is allowed during the evaluation phase.
    % \item \textbf{Independent evaluation team}. 
\end{itemize}

% \end{document}
% \subsection{Performance Metrics}

% Quantitative.

% \subsection{Evaluation Protocol}

% How to ensure fair evaluation (same platform, same setting, evaluators

\section{Performance}

% \documentclass[../main.tex]{subfiles}
% \graphicspath{{\subfix{../fig/}}}

% \begin{document}
% daily sharpe ratio is impossible larger than 9.00

\subsection{Performance Metrics}

We use the well-established quantitative metrics~\cite{liu2021finrl} in finance to evaluate models, as shown in Table~\ref{tab:metrics}. Specifically, we include cumulative and annualized returns to measure overall profitability and incorporate risk-adjusted performance metrics such as the Sharpe, Sortino, and Calmar ratios. Additionally, we consider downside risk indicators, including annualized volatility and maximum drawdown. For each metric, we report its definition, numerical range, and optimization direction (i.e., whether higher or lower values indicate better performance).

\begin{table}[]
    \centering
    \begin{tabular}{c p{6cm} c c}
    \toprule
         \textbf{Metric} &  \textbf{Definition} & \textbf{Range} & \textbf{Direction}\\
    \midrule
         Cumulative return & The total return generated by the trading strategy over a trading period. & $(-\infty, +\infty)$ & Higher is better\\
    \midrule
         Annualized return & The geometric average amount of money earned by the agent each year over a given time period. & $(-\infty, +\infty)$ &  Higher is better\\
    \midrule
         Annualized volatility & The annualized standard deviation of daily returns. & $[0, +\infty)$ & Lower is better\\
    \midrule
         Sharpe ratio & The excess return per unit of volatility. & $(-\infty, +\infty)$ & Higher is better\\
    \midrule 
          Maximum drawdown & The largest single drop in the portfolio value from peak to trough.& $[-1,0]$& Higher is better\\
    \midrule
          Rachev Ratio & A risk-adjusted return that measures the potential upside gain compared to potential downside risk, using the tails of the return distribution. &$(-\infty, +\infty)$ & Higher is better\\
    \midrule
          Return over maximum drawdown (RoMaD) & It is calculated as the cumulative return divided by the absolute value of maximum drawdown. &$(-\infty, +\infty)$ & Higher is better\\
    \midrule  
          Sortino ratio & The excess return divided by the downside deviation.& $(-\infty, +\infty)$& Higher is better\\
    \midrule 
          Calmar ratio & The annualized excess return divided by the maximum drawdown.& $(-\infty, +\infty)$& Higher is better\\
    \midrule  
          Omega ratio & It compares the probability of achieving returns above a threshold to the probability of falling below it. & $[0, +\infty)$& Higher is better\\
    \midrule 
          Win/Loss ratio & It is calculated by dividing the number of winning trades by the number of losing trades. &$[0, +\infty)$ & Higher is better\\
    \bottomrule
    \end{tabular}
    \caption{Evaluation metrics.}
    \label{tab:metrics}
\end{table}

\subsection{Stock Trading Tasks}

\subsubsection{Data-Centric Stock Trading}
\begin{table}[]
    \centering
    \begin{tabular}{c c c c c}
        \toprule
        \textbf{Team} & \textbf{Testing Period} & \textbf{Cumulative Return} & \textbf{Sharpe Ratio$^*$} & \textbf{Maximum Drawdown} \\
        \midrule
        \multirow{2}{*}{SZU-FIN-621}
              & 10/25/2023-11/14/2023 & 1.05\% & 2.26 & -1.36\%\\
               & 11/15/2023-11/22/2023 & -0.03\% & -10.25 & -0.03\%\\
        \midrule
        \multirow{2}{*}{Nik-Elena} 
            & 10/25/2023-11/14/2023 & 3.50\% &  9.56 & -0.40\%\\
            & 11/15/2023-11/22/2023 & 0.04\% &  0.95 & -0.15\%\\
        \midrule
        \multirow{2}{*}{WeCan} 
            & 10/25/2023-11/14/2023 & 2.35\% &  8.07 & -0.55\%\\
            & 11/15/2023-11/22/2023 & -0.05\% &  -1.74 & -0.11\%\\
        \bottomrule
        \multirow{2}{*}{DJIA} 
            & 10/25/2023-11/14/2023 & 5.42\% & 0.45  & -1.87\%\\
            & 11/15/2023-11/22/2023 & 0.80\% & 0.49  & -0.18\%\\    
        \bottomrule
    \end{tabular}
    \caption{The performance of top winner teams for FinRL Contest 2023 Task 1 Data-Centric Stock Trading. The stocks are constituents of the Dow Jones Index $(K=30)$. $^*$The results of the Sharpe ratio were reported wrongly.}
    \label{tab:stock_trading_performance}
\end{table}

The data-centric stock trading task in FinRL Contest 2023 is to train a trading agent by using novel data and feature engineering strategies. The training dataset is the OHLCV dataset with 10 market indicators for all 30 constituent stocks in the Dow Jones Index from 07/01/2010 to 10/24/2023 ($3,352$ trading days). The market indicators are shown in Table \ref{tab:technical_indicators}. The models are evaluated for two periods: 10/25/2023-11/14/2023 (15 trading days, pre-submission deadline) and 11/15/2023-11/12-2023 (6 trading days, post-submission deadline). Table \ref{tab:stock_trading_performance} shows the cumulative return, Sharpe ratio, and maximum drawdown of the top three winners and the Dow Jones Index. The teams used different approaches \footnote{GitHub repo for models and reports: \url{https://github.com/Open-Finance-Lab/FinRL_Contest_2023/tree/main/Task_1}}:
\begin{itemize}
    \item \textbf{SZU-Fin-621} employed PPO-Switch, which is an ensemble method combining multiple PPO agents. The agent pool was created by training different PPO agents with diverse online portfolio selection (OPS) \cite{yin2021ops} price features. The agent was selected based on its short-term and long-term returns, and the action space was sparsified to enhance capital efficiency \cite{philipp2020sparse}.
    \item \textbf{Nik-Elena} added new technical indicators in the state, including the Relative Strength Index for different periods, On-Balance Volume, and Moving Average for different periods. The team also enlarged the negative return when the turbulence exceeded a threshold.
    \item \textbf{WeCan} added 101 formulaic alpha factors \cite{alpha101} as new features in the state, in addition to the provided market indicators.
\end{itemize}

The results show that the teams' strategies achieved better risk-adjusted returns and risk management, but showed poorer profitability compared with the Dow Jones Index. In the first testing period, from 10/25/2023 to 11/14/2023, all teams outperformed the Dow Jones Index in the Sharpe ratio and maximum drawdown. Nik-Elena achieved the highest Sharpe ratio of 9.56  and the best maximum drawdown of -0.40\%, indicating excellent risk-adjusted performance and effective downside protection. In the second testing period from 11/15/2023 to 11/22/2023, SZU-FIN-621 and WeCan had negative cumulative returns and Sharpe ratios. Nik-Elena had a small positive return of 0.04\% but still underperformed the market index. However, Nik-Elena still achieves a higher Sharpe ratio and maximum drawdown than the Dow Jones Index, indicating better risk control even in challenging market environments. Participants' models showed strong risk management compared to the DJIA in the first testing period, but their generalization to new, unseen market conditions remains a challenge.

\subsubsection{Stock Trading with Ensemble Methods}

\begin{table}[]
  \centering
  \begin{tabular}{lcccccccc}
    \toprule
    Model & Ensemble-1 & Ensemble-2 & Ensemble-3 & PPO & SAC & DDPG & Min-Variance & DJIA \\
    \midrule
    Cumulative Return & 62.60\% & 58.77\% & 46.89\% & \textbf{63.37}\% & 50.62\%  & 63.19\% & 13.9\% & 18.95\% \\
    Annual Return & 18.22\% & 17.25\% & 14.15\% & \textbf{18.41} \% & 15.14\% & 18.36\% & 7.34\% & 6.15\% \\
    Annual Volatility & 11.76\% & 12.61\% & 12.70\% & \textbf{11.35}\% & 11.67\% & 11.93\% & 18.16\% & 15.14\%\\
    Sharpe Ratio & 1.48 & 1.33 & 1.11 & \textbf{1.55} & 1.27 & 1.47 & 0.48 & 0.47 \\
    Sortino Ratio & 2.34 & 2.14 & 1.74 & \textbf{2.44} & 2.05 & 2.37 & 0.73 & 0.67\\
    Max Drawdown & \textbf{-8.98\%} & -11.27\% & -12.27\%  & -9.96\%  & -12.02\%  & -13.15\% & -14.9\% & -21.94\% \\
    RoMaD & \textbf{6.97} & 5.22 & 3.82 & 6.36 & 4.21 & 4.81 & 1.10 & 0.86 \\
    Calmar Ratio & \textbf{2.03} & 1.53 & 1.15 & 1.85 & 1.26 & 1.40 & 0.49 & 0.28 \\
    Omega Ratio & 1.31 & 1.28 & 1.23 & \textbf{1.33} & 1.27 & 1.32 & 1.09 & 1.08 \\
  \bottomrule
\end{tabular}
\caption{Stock trading task performance. Models are trained, validated, and tested on a rolling window basis on OHLCV datasets for 30 Dow Jones stocks. Ensemble models use weighted averages on agent action probabilities.}
\label{tab:ensemble_performance_stock}
\end{table}

We performed the stock trading task for 30 stocks in the Dow Jones index by using three ensemble models, and individual PPO, SAC, and DDPG agents. 

\textbf{Stock data}: We use historical daily OHLCV data for all 30 stocks in the Dow Jones index from 01/01/2021 to 12/01/2023 (734 trading days). OHLCV data is a rich source for learning financial market behaviors and trends. We use technical indicators listed in Table \ref{tab:technical_indicators}. These indicators enrich the data with more insights into market behaviors and trends. Therefore, $K=30$ and $I=10$ in the setting of Section \ref{sec:mdp}.

\textbf{Agents for stock trading task}. We use PPO, SAC, and DDPG agents. The policy network for each agent consists of a feed-forward network with two hidden layers, having 64 units and 32 units, respectively. We set a learning rate of $3\cdot10^{-4}$ and a batch size of $64$. All ensemble models and individual agents are trained, validated, and tested on \textbf{a rolling-window basis} with 30-day training, 5-day validation, and 5-day testing windows.

\textbf{Ensemble methods}. The first method (Ensemble 1) consists of $1$ PPO, $1$ SAC, and $1$ DDPG agents; the second method (Ensemble 2) consists of $5$ PPO, $5$ SAC, and $5$ DDPG agents; and the third method (Ensemble 3) consists of $10$ agents for each type. As in Section \ref{sec:stock_ensemble}, all three ensemble models use the weighted average approach to combine component agent action probabilities. 

\begin{figure}[t]
  \centering
  \includegraphics[width=\linewidth]{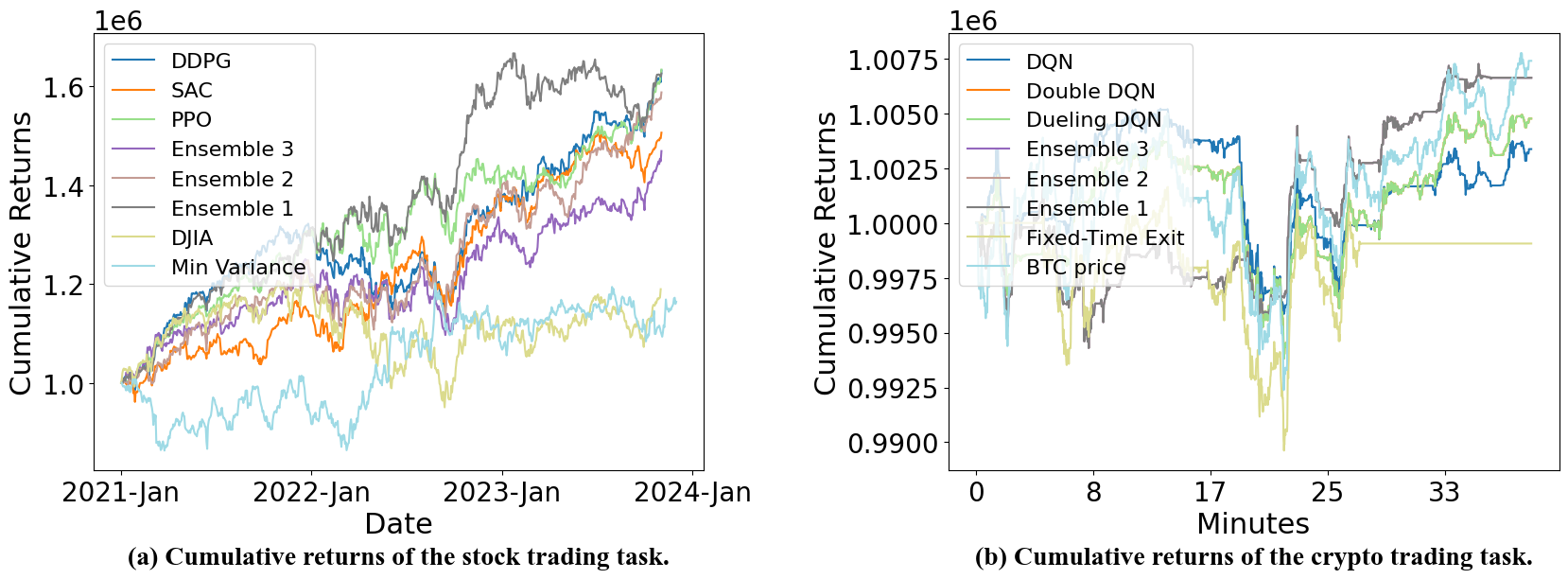}
  \caption{Cumulative returns of different strategies for the stock trading task and crypto trading task.}
  \label{fig:stock_returns}
\end{figure}

\textbf{Results}.  As seen in Table \ref{tab:ensemble_performance_stock}, the PPO agent achieves the highest cumulative returns of $63.37\%$, Sharpe ratio of $1.55$, and Sortino ratio of $2.44$, showing an ability to maintain high returns with controlled volatility and downside risk. Although DDPG’s cumulative returns are comparable to PPO's, its higher maximum drawdown of $-13.15\%$ signals a greater risk of large value drops, which is a concern for risk management. SAC has a lower maximum drawdown than DDPG but underperforms in other metrics. All individual agents significantly outperform two traditional baselines across all metrics. The ensemble models also maintain profitability and risk management advantages over the baselines. Ensemble 1 has a high cumulative return of $62.60\%$, and as shown in Fig. \ref{fig:stock_returns} (a), it shows superior performance from Sep 2022 to Oct 2023. Ensemble 1 also achieves the smallest maximum drawdown and a higher Sharpe ratio than SAC and DDPG. Ensembles 1 and 2 have high RoMaD and Calmar ratios, showing an ability to quickly recover from peak-to-trough losses and a potential for steady growth in market adversities.

\subsubsection{LLM-engineered Signals for Stock Trading}

% \begin{table}[]
%     \centering
%     \begin{tabular}{c c c c}
%     % \toprule
%     %      \textbf{Team} & \textbf{Task} & \textbf{Cumulative Return}	& \textbf{Win Rate} & \textbf{Loss Rate}\\
%     % \midrule
%     %      aethernet & LLM-Engineered Signals with RLMF & 1.34\% & 0.60 & 0.40 \\
%     % \bottomrule
%     \toprule
%         \textbf{Team} &  \textbf{Cumulative Return}	& \textbf{Rachev Ratio} & \textbf{Maximum Drawdown}\\
%     \midrule
%          Ruijian \& Sally & 335.58\% & 1.09 & -50.24\% \\
%     \bottomrule
%         Nasdaq-100 & 164.52\% & 0.94 & -35.56\%\\
%         S\&P500 & 90.03\% & 0.90 & -33.93\%\\
%     \bottomrule
%     \end{tabular}
%     \caption{The performance of top winner teams for FinRL Contest 2025 Task 1 FinRL-DeepSeek for Stock Trading.}
%     \label{tab:stock_trading_llm_performance}
% \end{table}
We designed two tasks around LLM-engineered signals from news, including FinRL Contest 2024 LLM-Engineered Signals with RLMF and FinRL Contest 2025 FinRL-DeepSeek for Stock Trading.

\textbf{FinRL Contest 2024 LLM-Engineered Signals with RLMF}. Participants fine-tuned LLMs to generate signals through reinforcement learning from market feedback. The training dataset was daily OHLCV data and financial news for $K=7$ large-cap stocks from 01/01/2020 to 10/08/2022 (447 trading days). The effectiveness of LLM-engineered signals is evaluated on a dataset from 10/09/2022 to 12/15/2023 (229 trading days), using a long-short trading strategy, i.e., long (buy) the three stocks with the highest sentiment scores and short (sell) the bottom three, closing all positions three days later. \textbf{Aethernet}\footnote{\url{https://github.com/Arnav-Gr0ver/ICAIF_FinRL-2024}}\cite{grover2025finrllama} fine-tuned the LLaMA-3.2-3B-Instruct model through RLMF to generate sentiment scores from news. The team designed a reward function to assess the sentiment score based on market data. The model had a cumulative return of 134.05\%, while a simple buy-and-hold strategy yields a 72.71\% return. It indicates that the LLM-engineered signals were effective even using a simple long-short execution strategy. The win/loss ratio of 1.5 suggests that the model generated more profitable trades than losing ones.

\begin{table}[t]
\centering
\begin{tabular}{ l l c c c c}
\toprule
\textbf{Team} & \textbf{Algorithm} & \textbf{Cumulative Return} & \textbf{Sharpe Ratio} & \textbf{Max Drawdown} & \textbf{Rachev Ratio} \\
\midrule
Otago Alpha & PPO & 191.14\% & 1.0800 & -28.16\% & 1.0557 \\
\midrule
Ruijian \& Sally & Refined GRPO  & 335.57\% & 0.9500 & -50.24\% & 1.0300 \\
\midrule
\multirow{6}{*}{Kuxlnw} 
 & Adaptive Portfolio & 163.88\% & 0.0574 & -33.10\% & 0.9389 \\
 & PPO 100 & 204.51\% & 0.0671 & -36.56\% & 1.0027 \\
 & CPPO 100 & 90.52\% & 0.0359 & -34.78\% & 1.0194 \\
 & PPO-DeepSeek 100 & 94.43\% & 0.0433 & -34.09\% & 0.9441 \\
 % & Mini CPPO & 13.95\% & 0.0147 & -42.29\% & 0.9266 \\
 % & Mini PPO & 130.59\% & 0.0527 & -34.23\% & 0.8790 \\
 & CPPO-DeepSeek 100 & 91.26\% & 0.0383 & -43.03\% & 0.8684 \\
\midrule
\multirow{2}{*}{Queen's Gambit} 
 & PPO with MIST 1 & 238.70\% & 0.2859 & -92.20\% & 6.7999$^*$ \\
 & PPO with MIST 2 & 342.65\% & 0.2897 & -92.47\% & 6.6723$^*$ \\
\bottomrule
S\&P 500 & / & 90.03\% & 0.0448 & -33.93\% & 0.9047 \\
Nasdaq-100 & / & 164.52\% & 0.0558 & -35.56\% & 0.9434 \\
\bottomrule
\end{tabular}
\caption{The performance of top winner teams for FinRL Contest 2025 Task 1 FinRL-DeepSeek for Stock Trading. $^*$ The Rachev ratio of Queen's Gambit is reported wrongly and needs further scrutiny.}
\label{tab:stock_trading_llm_performance}
\end{table}

\textbf{FinRL Contest 2025 FinRL-DeepSeek for Stock Trading}. This task uses a different approach to combine FinRL and LLM-engineered signals, where the signals are integrated into the environment to train FinRL agents. The training dataset is daily OHLCV data with 10 market indicators (listed in Table \ref{tab:technical_indicators}) and financial news for Nasdaq 100 constituent stocks from 01/01/2013 to 12/31/2018 (1510 trading days). The models are evaluated on the testing dataset from 01/01/2019 to 12/31/2023 (1258 trading days). The winning teams used different approaches:
\begin{itemize}
    % \item \textbf{Aethernet}\cite{grover2025finrllama}\footnote{\url{https://github.com/Arnav-Gr0ver/ICAIF_FinRL-2024}} fine-tuned the LLaMA-3.2-3B-Instruct model through RLMF to generate sentiment scores from news. The training dataset was daily OHLCV data and financial news for $K=7$ large-cap stocks from 01/01/2020 to 10/08/2022 (447 trading days). The team designed a reward function to assess the sentiment score based on market data. The effectiveness of sentiment scores is evaluated on a dataset from 10/09/2022 to 12/15/2023 (229 trading days), using a long-short trading strategy, i.e., long (buy) the three stocks with the highest sentiment scores and short (sell) the bottom three, closing all positions three days later. The model had a cumulative return of 134.05\%, while a simple buy-and-hold strategy yields a 72.71\% return. It indicates that the LLM-engineered signals were effective even using a simple long-short execution strategy. The win/loss ratio of 1.5 suggests that the model generated more profitable trades than losing ones. 
    \item \textbf{Otago Alpha} \cite{li2025otago} added the put-call ratio to the feature vector, which is an assessment of put and call option transaction volumes. The put-call ratio was sourced from the OptionMetrics via the Wharton Research Data Services (WRDS) database and Bloomberg Terminal. %Their PPO agent achieved a cumulative return of 191.14\%, outperforming both the Nasdaq-100 (164.52\%) and the S\&P 500 (90.03\%). The Rachev ratio of 1.06 indicates that the strategy generated greater upside gains than downside losses during extreme market conditions. It also achieved a maximum drawdown of -28.16\%, indicating lower downside risk, compared with Nasdaq-100 (-35.57\%) and the S\&P 500 (-33.93\%).
    \item \textbf{Ruijian \& Sally}\footnote{\url{https://github.com/Ruijian-Zha/FinRL-DAPO-SR}}\cite{zha2025ruijian} used the Group Relative Policy Optimization (GRPO) \cite{yu2025dapo} algorithm, which is refined by Decoupled Clipping and Dynamic sAmpling Policy Optimization (DAPO). The team also designed a reward function adjusted by sentiment and risk. %The model is evaluated on the testing dataset from 01/01/2019 to 12/31/2023 (1258 trading days). It achieved a cumulative return of 335.58\%, outperforming both the Nasdaq-100 (164.52\%) and the S\&P 500 (90.03\%). The Rachev ratio of 1.09 indicates that the strategy generated greater upside gains than downside losses during extreme market conditions. However, the team has a maximum drawdown of -50.24\%, indicating a large downside risk. 
    \item \textbf{Kuxlnw}\footnote{\url{https://github.com/Vorakorn1001/FinRL-DeepSeek}}\cite{vorakorn2025kuxlnw} adopted a dynamic model selection mechanism that adapts to market conditions and macroeconomic indicators. The team added new features to the environment, including interest rates, gold prices, and oil prices. They also used DeepSeek-generated sentiment scores and risk levels in the features. They trained specialized FinRL agents for each stock. Then they employed a mechanism to dynamically allocate the most suitable trading agents based on bull or bear markets. The algorithms they use include PPO with 100 training epochs (PPO 100), PPO with DeepSeek-generated signals and 100 training epochs (PPO-DeepSeek 100), CPPO with 100 training epochs (CPPO 100), CPPO with DeepSeek-generated signals and 100 training epochs (CPPO-DeepSeek 100), and Adaptive Portfolio applying the dynamic model selection mechanism.
    \item \textbf{Queen's Gambit}\footnote{\url{https://github.com/sahar-arshad/QueensGambit-FinRL2025}}\cite{arshad2025queen} proposed a two-phase framework, Market-Informed Sentiment for Trading (MIST), to extract signals from news. In the first phase, it used a structured prompt with few-shot examples for DeepSeek-R1 7B to evaluate financial news and assign a sentiment score (1–5). In the second stage, the team designed an advanced prompt incorporating market behavior (i.e., short-term stock price change direction) to either reinforce or contradict the sentiment extracted in the first stage. The final trading signal is integrated into the environment to train FinRL agents. The team trained two PPO agents, one with only the first stage of MIST (PPO with MIST 1) and one with two stages of MIST (PPO with MIST 2).
\end{itemize}

The performance of the winning teams are shown in Table \ref{tab:stock_trading_llm_performance}. Team Otago Alpha, Ruijian \& Sally, and Queen’s Gambit showed a superior performance over S\&P 500 and Nasdaq-100 in cumulative return and Sharpe ratio. Otago Alpha achieved the highest Sharpe ratio ($1.08$) and maximum drawdown ($-28/16\%$), and the cumulative return outperformed two market indices, showing a good balance between profitability and risk management. Although Ruijian \& Sally and Queen’s Gambit showed high cumulative returns, they had a worse maximum drawdown of $-50.24\%$ and $-92.47\%$ respectively compared with two market indices, indicating a larger downside risk. A Rachev ratio larger than 1 indicates that their strategies' upside tail rewards outweighed their downside tail risks during extreme market movements. Kuxlnw's adaptive portfolio underperformed its PPO 100 model in cumulative return, Sharpe ratio, and Rachev ratio, suggesting that a multi-asset trading agent may have more stable performance in this setting.

\subsection{Crypto Trading Tasks}

\begin{table}[t]
\centering
  \begin{tabular}{lcccccccc}  
    \toprule
    Model           & Ensemble-1            & Ensemble-2            & Ensemble-3        & DQN     & Double DQN   & Dueling DQN & Fixed-Time Exit & BTC Price \\
    \midrule
    Cumulative Return   & 0.66\%            & 0.66\%                & 0.66\%            & 0.34\%    & 0.48\%    & 0.48\%    & -0.1\%    & \textbf{0.74\%} \\
    Sharpe Ratio        & \textbf{0.28}      & \textbf{0.28}       & \textbf{0.28}      & 0.15      & 0.21     & 0.21      & -0.03     & 0.20 \\
    Maximum Drawdown        & \textbf{-0.73\%}   & \textbf{-0.73\%}   & \textbf{-0.73\%}   & -0.93\%   & -0.98\%   & -0.98\%   & -1.00\%   & -1.3\%  \\
    RoMaD               & \textbf{0.90}     & \textbf{0.90}      & \textbf{0.90}      & 0.37      & 0.49      & 0.49      & 0.10      & 0.59 \\
    Sortino Ratio       & \textbf{0.39}      & \textbf{0.39}      & \textbf{0.39}      & 0.20      & 0.29     & 0.29      & -0.04     & 0.28 \\
    Omega Ratio         & \textbf{1.08}      & \textbf{1.08}      & \textbf{1.08}       & 1.04      & 1.05      & 1.05      & 0.99      & 1.05\\
    %Win Rate            & \textbf{61.86\%}   & \textbf{61.86\%}   & \textbf{61.86\%}   & 56.70\%   & 61.79\%   & 61.79\%   & 4.00\%            & - \\
    %Loss Rate           & 38.14\%           & 38.14\%               & 38.14\%            & 43.30\%          & 38.21\%  & 38.21\%   & \textbf{5.10\%}    & - \\
    Win/Loss Ratio & \textbf{1.622}  & \textbf{1.622}  & \textbf{1.622} & 1.309 & 1.617 & 1.617 & 0.5 & - \\
  \bottomrule
\end{tabular}
\caption{crypto trading task performance. The second-level LOB data for Bitcoin is split into out-of-sample data for training and in-sample data for testing. Ensemble models use majority voting on agent actions.}
\label{tab:ensemble_performance_crypto}
\end{table}

The FinRL Contest 2024 crypto Trading with Ensemble Learning and the FinRL Contest 2025 FinRL-AlphaSeek for Crypto Trading are tasks designed for Bitcoin trading.

\subsubsection{Crypto Trading with Ensemble Methods}
In our experiment, the crypto trading task for Bitcoin (BTC) $(K = 1)$ is performed using three ensemble models, and individual DQN, Double DQN, and Dueling DQN agents.

\textbf{crypto data}: The dataset comprises second-level LOB data for BTC from 04/07/2021 11:32:42 to 04/19/2021 09:54:22. Adaptations from 101 formulaic alphas \cite{alpha101} are calculated based on LOB data to extract insights into market behaviors, such as momentum, mean-reversion, and market anomalies. A recurrent neural network (RNN) further processes the 101 alphas into $I=8$ technical indicators. It reduces the complexity of the input data and enhances the ability to predict market trends, thus improving generalization and avoiding overfitting. The RNN is trained on data from 04/07/2021 11:32:42 to 04/17/2021 00:38:02 without future information leaks. The agents are trained on the in-sample data from 04/17 00:38:03 to 04/19 09:09:21 and tested on the out-of-sample data from 04/19 09:09:22 to 04/19 09:54:22.

\textbf{Agents for crypto trading task}.
%the RNN uses parallel LSTM and GRU layers and combines their outputs when processing 101 alphas. 
We use DQN, Double DQN, and Dueling DQN agents. The policy network for each agent consists of a feed-forward neural network with three 128-unit hidden layers. We set an exploration rate of $0.005$, a learning rate of $2\cdot 10^{-6}$, and a batch size of $512$. 

\textbf{\textbf{Ensemble methods}} The first method (Ensemble 1) consists of $1$ DQN, $1$ Double DQN, and $1$ Dueling DQN agents; the second method (Ensemble 2) consists of $3$ DQN, $3$ Double DQN, and $3$ Dueling DQN agents; and the third method (Ensemble 3) consists of $10$ agents for each type. As in Section \ref{sec:crypto_ensemble}, three ensemble models use a majority voting approach to aggregate the agents' actions. 
All ensemble models and individual agents are trained on the in-sample data and tested on the out-of-sample data. 

\textbf{Results}. As seen in Table \ref{tab:ensemble_performance_crypto} and Fig. \ref{fig:stock_returns} (b), Double DQN and Dueling DQN agents have similar performance, with cumulative returns of $0.48\%$. This is lower than the BTC price baseline. Despite this, they achieve higher Sharpe ratios of $0.21$ and lower maximum drawdowns of $-0.98\%$ than the fixed-time exit strategy and BTC price baseline, suggesting effective risk management. The three different ensemble models have similar performances, which may be due to the limited action space at each timestep, causing agents to output identical actions. Their cumulative returns are close to the BTC price baseline. Moreover, the ensemble models outperform all individual agents in all metrics, achieving the highest Sharpe ratio of $0.28$ and the lowest maximum drawdown of $-0.73\%$. They also have the highest win/loss ratio of $1.62$. This shows that ensemble methods can mitigate the risks associated with the decision-making failures of single agents.

We observe that individual agents and majority voting ensembles have near identical performances; in the restricted action space, agent policies may be converging, indicating a greater risk of significant losses due to less diversified trading actions. Compared to the spot price and fixed-time exit, the ensemble consistently achieved a higher return over the maximum drawdown ratio, highlighting superior risk-adjusted returns. Due to a lack of diversity in agents, we observed near-identical results between ensembles and individual agents. Nonetheless, over a relatively short timespan of $30$ minutes, we observe that the vectorized agents can outperform other strategies.

This was crucial in maintaining portfolio stability amid high market volatility. The ensemble generally showed a higher win rate and a lower loss rate than the individual agents, reflecting its enhanced decision-making accuracy and consistency. The comparative analysis between the ensemble model and individual trading agents underscores the ensemble’s superior capability in managing risks and capitalizing on market opportunities. While individual agents provide valuable insights and are crucial components of the ensemble, the aggregated approach of the ensemble offers a more robust and effective solution for trading in the volatile crypto markets. 

\subsubsection{Crypto Trading in FinRL Contests}

\begin{table}[]
    \centering
    \begin{tabular}{c c c c c}
    \toprule
        \textbf{Contest} & \textbf{Team} &  \textbf{Cumulative Return}	& \textbf{Sharpe Ratio} & \textbf{Maximum Drawdown}\\
    \midrule
         \multirow{1}{*}{FinRL Contest 2024} 
            & Fermion & 0.23\% & 0.0037 & -0.28\% \\
            % & RL3 & -0.80\% & 0.0084 & -10.6\% \\
    \midrule
         \multirow{1}{*}{FinRL Contest 2025} 
            & Mt.Everest  & -0.05\% & -0.0008 & -1.23\% \\
            % & cryptoalphaseek & -4.67\% & -0.03 & -5.54\% \\ 
    \bottomrule
        Baseline & BTC price & -7.35\% & -0.0022 & -17.02\%\\
    \bottomrule
    \end{tabular}
    \caption{The performance of winners for FinRL Contest 2024 Task 1 ``crypto Trading with Ensemble Learning'' and FinRL Contest 2025 Task 2 ``FinRL-AlphaSeek for Crypto Trading''.}
    \label{tab:crypto_trading_performance}
\end{table}

Based on the previous experiments, we designed tasks allowing participants to explore novel factor mining strategies or ensemble methods. The training dataset is a second-level LOB dataset for Bitcoin from 04/07/2021 11:32:42 to 04/17/2021 00:38:02. The testing dataset is from 04/17/2021 00:38:02 to 04/19/2021 09:54:22. We re-partitioned the dataset to allow for a longer testing period. The performance of winners is shown in Table \ref{tab:crypto_trading_performance}. 
\begin{itemize}
    \item \textbf{Fermion} used ensemble methods to combine the strengths of different on-policy and off-policy algorithms. The team applied majority voting based on confidence scores to determine the trading action. 
    \item \textbf{Mt.Everest} added new technique indicators in the state of the market environment, including order book imbalance, spread-based liquidity, order flow imbalance, price impact of trades, market-to-limit order ratio, intraday volatility, momentum scores, and noise-to-signal ratio. In addition, the team used principal component analysis (PCA) to extract the top 10 strongest factors from 101 alpha factors \cite{alpha101}. To construct the agent pool, they applied different policy networks, including LSTM, transformer, and dilated CNNs. The action was determined through majority voting.
\end{itemize}

Both teams outperformed the baseline BTC price in cumulative returns, Sharpe ratio, and maximum drawdown. Fermion achieved a positive cumulative return of 0.23\% and a Sharpe ratio of 0.0037. Both teams' higher maximum drawdown shows better downside risk control and loss mitigation, which is a notable achievement given BTC's inherent volatility. 

In addition, FinRL-Crypto\footnote{\url{https://github.com/AI4Finance-Foundation/FinRL_Crypto}} \cite{gort2022deep} introduces a multi-crypto trading strategy. The agents were trained using five-minute-level OHLCV data for $K=10$ cryptocurrencies from 02/02/2022 to 04/30/2022. The agents were trained with multiple hyperparameter settings across 10 training-validation splits. To address backtest overfitting, agents with a high probability of overfitting were rejected at a $10\%$ significance level. The model was evaluated on the dataset from 05/01/2022 to 06/27/2022, which includes two market crashes. The best-performing PPO agent achieved a cumulative return of $-34.96\%$, outperforming the S\&P crypto Broad Digital Market Index (S\&P BDM Index, $-50.78\%$) and the equal-weight strategy ($-47.78\%$). It also had lower volatility ($0.0020$), compared with S\&P BDM Index ($0.0581$) and the equal-weight strategy ($0.0042$).

% \end{document}

\section{Organizational Aspects}
% \documentclass[../main.tex]{subfiles}
% \graphicspath{{\subfix{../fig/}}}
% \begin{document}

% 7 contients.
% correct statistics

We held FinRL contests at several academic conferences, including ACM International Conference on AI in Finance (ACM ICAIF 2023, 2024), IEEE International Conference on Intelligent Data and Security (IEEE IDS 2025), and IEEE International Conference on Cyber Security and Cloud Computing (IEEE CSCloud 2025). In this section, we summarize the organizational experience of the FinRL Contests, including encouraging participation, contest process, platform setup, event promotion, registration and submission approaches, and communication. 

\begin{figure}
    \centering
    \includegraphics[width=\linewidth]{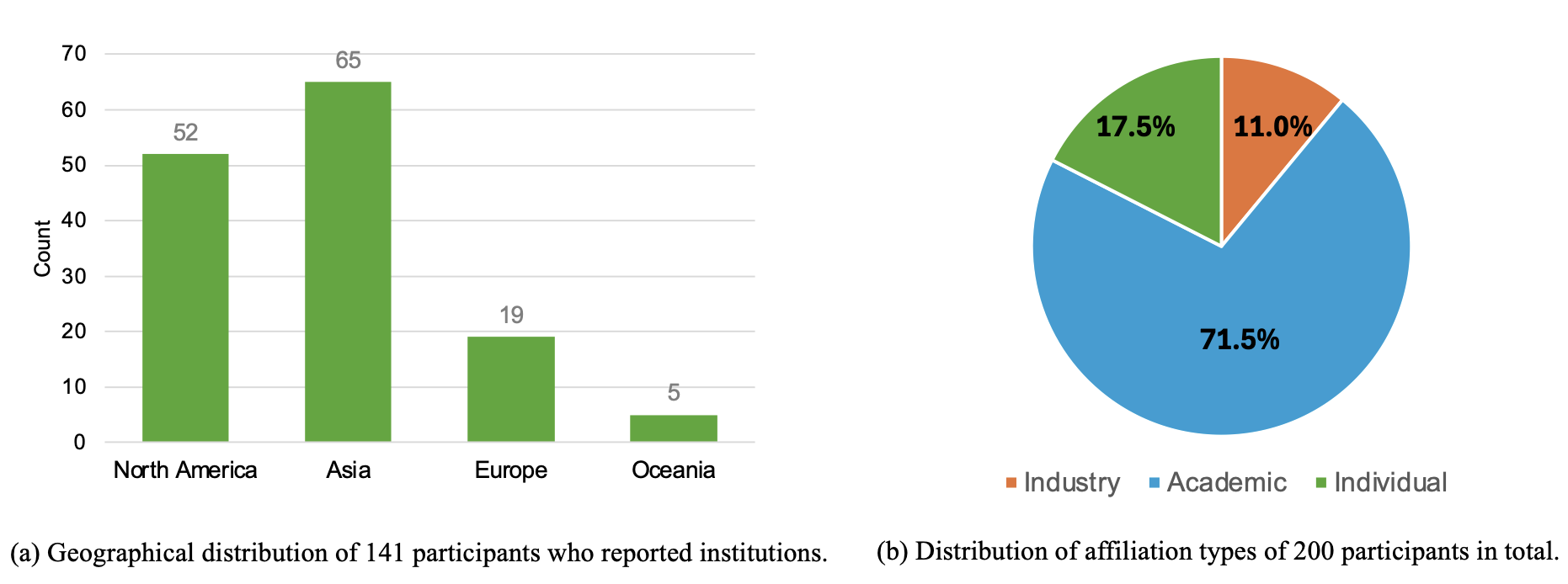}
    \caption{Distribution of participation by geography and affiliation type.}
    \label{fig:participants}
\end{figure}

\textbf{Participation}. From 2023 to 2025, a total of 200+ students, researchers, and practitioners participated. Fig. \ref{fig:participants} shows the distribution of participation by geography and type of affiliation. Among the 141 participants who reported their institutions, $37\%$ of participants are from North America, including the United States and Canada. $46\%$ of participants come from Asia, mainly from China, South Korea, and India. Among the 200+ participants, $71.5\%$ of them come from 62 academic institutions, such as Columbia University, the National University of Singapore, and Shenzhen University. $17.5\%$ of them are from 21 industrial institutions. (The statistics exclude the ongoing FinAI Contest 2025 @ IEEE CSCloud).

\textbf{Contest Process}. We first set important dates, including team registration, starter-kit release, model submission deadline, report submission deadline, and leaderboard release. The full contest typically lasts 2–3 months. It aims to provide participants with sufficient time for solution development while ensuring timely evaluation and feedback.
\begin{itemize}
    \item \textbf{Team registration}. Each team consists of 1-4 people. Team registration starts as early as possible to leave enough time for promotion. The starter kit is nearly ready when team registration begins so that registered teams can be engaged and start early. Registration remains open until the model submission deadline to continuously encourage new teams to join.
    \item \textbf{Starter kit release}. The starter kit is released within a week after team registration begins. This short buffer period is used to organize and clean the prepared datasets, code, and instructions. The early release ensures that teams can be engaged early and have ample time to develop their solutions.
    \item \textbf{Model submission}. The model submission deadline is set more than a month after the starter kit release. This period gives teams sufficient time to get familiar with the tasks, study the tutorials, explore the dataset, use the code, and develop their own models.
    \item \textbf{Report submission}. In addition to models, participant teams are encouraged to submit a 2-3 page short paper describing the methodologies and performance. The report submission deadline is set one week after the model submission deadline. Participant teams can finalize documentation after completing model development. We invited a panel of 11 expert reviewers, including Ph.D. researchers, postdoctoral fellows, and professors from Columbia University, New York University Shanghai, Stevens Institute of Technology, and Princeton University. They review the submitted reports and give valuable revision advice.
    \item \textbf{Result announcement}. Final results are announced 15 days after the report submission deadline. During this period, evaluators assess the submitted models, and experts review the reports. Final rankings are determined based on a weighted score, 60\% of model performance and 40\% of the report assessment.
\end{itemize}

\textbf{Platform Setup}. Multiple platforms are set up to host different resources for the contest, as shown in Table \ref{tab:kits}:
\begin{itemize}
    \item \textbf{Contest website}. The website includes the contest overview, task description, contest timeline, registration and submission guidelines, contact information, and links to related resources. It is set up by using \textsf{GitHub Pages} for wide accessibility. It also serves as the main portal for announcements and updates.
    \item \textbf{GitHub repository}. A GitHub repo is set up to host the starter kit. The datasets, code, and detailed descriptions are organized in a folder for each task. It also contains around 8 tutorial notebooks for participant teams to get started.
    \item \textbf{Documentation website}. The documentation serves as a detailed guide and resource hub of FinRL Contests for participants. It includes FinRL introduction, task descriptions, and detailed explanations of the starter kit and baseline solutions.
\end{itemize}

\textbf{Event Promotion}. The FinRL Contests welcome students, researchers, and practitioners who are passionate about finance and machine learning. The contest is promoted widely through mailing lists of Google Groups and universities, and social media platforms, such as LinkedIn, Twitter, and Facebook, highlighting its inclusive and challenging nature. We also collaborate with some media partners, including Wilmott, PyQuant News, and Paris Machine Learning Group. 

\textbf{Registration and Submission}. Team registration and model submission are through \textsf{Google Forms} for easy management. Registration requires a team name and members’ information, including name, email, and affiliation. For model submission, teams provide a link to the GitHub and/or Hugging Face repository. The repository should be well organized, including code, model weights, scripts, and detailed instructions for evaluation. The report submission is through \textsf{OpenReview} or the channel required by the conference. 

\textbf{Communication Channel}. We have multiple channels to communicate with participants for their questions, announcements, and tutorials. 
\begin{itemize}
    \item \textbf{Contact email}. We use an official contact email, \textsf{finrlcontest@gmail.com}, to send important announcements and respond to inquiries.
    \item \textbf{Group chat}. We set up a Discord channel of 500+ people and 2 WeChat groups of 270+ people to facilitate participant networking, discussion, and direct Q\&A.
    \item \textbf{GitHub issue}. Participants are encouraged to create GitHub issues in the starter kit repository to seek technical support. 
    \item \textbf{Online sessions}. 3 online Q\&A sessions were organized via Zoom to address common concerns and questions. 1 online tutorial session was organized to introduce the FinRL framework, the contest, and the usage of the starter kit.
\end{itemize}

% \end{document}

% Conference acreddit
% Report judeges (professor, postdoc, ...)

\section{Conclusions and Future Works}\label{sec5}
In this paper, we present the financial reinforcement learning benchmark through FinRL Contests, held from 2023 to 2025. The contests provided standardized task definitions, curated market datasets, GPU-optimized parallel market environments, and defined evaluation metrics to ensure reproducibility and comparability across various financial applications such as stock trading, cryptocurrency trading, and the integration of signals from LLMs. Participant teams employed diverse strategies, including feature engineering, ensemble learning methods, and LLM-generated market signals, demonstrating significant progress in both performance and risk management compared to conventional market benchmarks. 

For future work, we will continue exploring and integrating LLM-generated signals from multimodal financial data in FinRL, such as SEC filings, earnings conference calls, alternative data, etc. The current breakthrough of LLMs in strong reasoning ability during inference time could also bring great potential in financial tasks. We will further develop FinRL trading agents that can be integrated into real-time intelligent trading systems, capable of processing and reasoning over multimodal data. We will utilize LLMs to process real-time financial data (e.g, \href{https://www.financialdatasets.ai/}{financial dataset} and \href{https://finnhub.io/docs/api/introduction}{Finnhub}) into trading signals, which will be put into the state of the trading agent. We will continue to actively integrate the most cutting-edge techniques with financial applications.

%\backmatter
\bmsection*{Author contributions}

% Add thanks to author contributors
This is an author contribution text. 
% This is an author contribution text. This is an author contribution text. This is an author contribution text. This is an author contribution text.

\bmsection*{Acknowledgments}

We thank Li Deng, Zihan Ding, Jian Guo, Zhouchi Lin, Christina Dan Wang, Zhaoran Wang, Matt White, Bo Wu, Xiaojun Wu, Zhuoran Yang, Daochen Zha, and Chuheng Zhang for serving as advisors of FinRL Contest 2023-2025. We thank Mostapha Benhenda, Jin Bo, Jiale Chen, Qian Chen, Arnav Grover, Ethan Havemann, Sarah Huang, Colin Lin, Shivan Mukherjee, Jaisal Patel, Charlie Shen, Kent Wu, Yangyang Yu, and Andy Zhu for serving as organizers, and Steve Ewald, Andrew Li, Nikola Maruszewski, Christopher Minn, and Gavin Wang for creating the evaluation platform.

We thank Zihan Dong, Allan Feng, Astarag Mohapatra, Louis Owen, Guoxuan Wang, Zhiyuan Wang, Bruce Yang, Luna Zhang, Jiahao Zheng, and Ming Zhu from the open-source AI4Finance community, who contributed to the open-source FinRL project. We also thank Jimin Huang from the FinAI community. We thank Renyuan Xu, Huining Yang, and Bo An for academic feedback on FinRL papers.

We thank all participant teams!

% \cite{Kenamond2013} Provide text here. This is acknowledgment text. Provide text here. This is acknowledgment text. Provide text here. This is acknowledgment text. Provide text here. This is acknowledgment text. Provide text here. This is acknowledgment text. Provide text here. This is acknowledgment text. Provide text here. This is acknowledgment text. Provide text here. This is acknowledgment text. Provide text here.

\bmsection*{Financial disclosure}

The authors thank Vatic Investment and the Shanghai Frontiers Science Center of Artificial Intelligence and Deep Learning at NYU Shanghai for covering registration fees for participant teams.

Keyi Wang, Nikolaus Holzer, Jiechao Gao, Anwar Walid, and Xiao-Yang Liu Yanglet acknowledge the support from Columbia's SIRS and STAR Program, as well as The Tang Family Fund for Research Innovations in FinTech, Engineering, and Business Operations. Xiao-Yang Liu Yanglet also acknowledges the support from a NSF IUCRC CRAFT center research grant (CRAFT Grant 22017) for this research. The opinions expressed in this publication do not necessarily represent the views of NSF IUCRC CRAFT.

\bmsection*{Conflict of interest}

The authors declare no potential conflict of interest.

\bibliography{wileyNJD-AMA}

% \bmsection*{Supporting information}

% Additional supporting information may be found in the
% online version of the article at the publisher’s website.

% \appendix
% % \subfile{sections/appendix}

% \bmsection*{Author Biography}

% \begin{biography}{\includegraphics[width=76pt,height=76pt,draft]{WileyNJDv5_Template/empty.pdf}}{
% {\textbf{Author Name.} Please check with the journal's author guidelines whether
% author biographies are required. They are usually only included for review-type articles, and typically require photos and brief biographies for each author.}}
% \end{biography}

\end{document}